\documentclass[twocolumn]{aastex631}

\usepackage{amsmath}
\usepackage{multirow}
\usepackage{graphicx}
\usepackage{booktabs}
\usepackage{color}

\newcommand{\halpha}{H$\alpha$}

\usepackage{scalerel}
\newcommand{\CaII}{{\ion{Ca}{2}}}

\newcommand{\CaIIk}{{\ion{Ca}{2}} K}

\newcommand{\mdot}{$\dot{\text{M}}$}
\newcommand{\Mdot}{{\dot{{M}}}}

\newcommand{\lsun}{ L_{\sun}}
\newcommand{\msunyr}{M_{\sun} \, \rm{ yr^{-1}}}
\newcommand{\kms}{ \, km \, s^{-1}}
\newcommand{\teff}{T$_{\rm eff}$}

\newcommand{\ri}{R$_{\rm i}$}

\newcommand{\Dr}{$\Delta \rm r$}
\newcommand{\tmax}{T$_{\rm max}$}

\received{June 14, 2024}
\revised{October 16, 2024}
\accepted{October 16, 2024}
\submitjournal{ApJ}

\shorttitle{}

\shortauthors{Micolta et al.}

\begin{document}

\title{
Using the Ca II lines in T Tauri stars to infer the abundance of refractory elements in the innermost disk regions
}

\correspondingauthor{Marbely Micolta}
\email{micoltam@umich.edu}

\author[0000-0001-8022-4378]{Marbely Micolta}
\affiliation{Department of Astronomy, University of Michigan, 1085 South University Avenue, Ann Arbor, MI 48109, USA}

\author[0000-0002-3950-5386]{Nuria Calvet}
\affiliation{Department of Astronomy, University of Michigan, 1085 South University Avenue, Ann Arbor, MI 48109, USA}

\author[0000-0003-4507-1710]{Thanawuth Thanathibodee}
\affiliation{Department of Physics, Faculty of Science, Chulalongkorn University, 254 Phayathai Road, Pathumwan, Bangkok, 10330 Thailand.}

\author[0000-0003-1166-5123]{Gladis Magris C.}
\affiliation{Centro de Investigaciones de Astronomía ``Francisco J. Duarte" CIDA, Av. Alberto Carnevali, Mérida 5101, Mérida, Venezuela}

\author[0000-0003-3562-262X]{Carlo F. Manara}
\affiliation{European Southern Observatory, Karl-Schwarzschild-Strasse 2, 85748 Garching bei München, Germany}

\author[0000-0002-4115-0318]{Laura Venuti}
\affiliation{SETI Institute, 339 Bernardo Ave., Suite 200, Mountain View, CA 94043, USA}

\author[0000-0001-8657-095X]{Juan Manuel Alcalá}
\affiliation{INAF–Osservatorio Astronomico di Capodimonte, Salita Moiariello 16, 80131 Napoli, Italy}

\author[0000-0002-7154-6065]{Gregory J. Herczeg}
\affiliation{Kavli Institute for Astronomy and Astrophysics, Peking University, Beijing 100871, People's Republic of China}

\begin{abstract}
We present a study of the abundance of calcium in the innermost disk of 70 T Tauri stars in the star-forming regions of Chamaeleon I, Lupus and Orion OB1b. 
We use calcium as a proxy for the refractory material that reaches the inner disk. 
We used magnetospheric accretion models to analyze the {\CaII} emission lines and estimate abundances in the accretion flows of the stars, which feed from the inner disks. We find Ca depletion in
disks of all three star-forming regions, with 57\% of the sample having $[Ca/H] < -0.30$ relative to the solar abundance. All disks with cavities and/or substructures show depletion, consistent with trapping of refractories in pressure bumps. Significant Ca depletion ($[Ca/H] < -0.30$) is also measured in 60\% of full disks, although some of those disks may have hidden substructures or cavities.
We find no correlation between Ca abundance and stellar or disk parameters except for the mass accretion rate onto the star. This could suggest that the inner and outer disks are decoupled, and that the mass accretion rate is related to a mass reservoir in the inner disk, while refractory depletion reflects phenomena in the outer disk related to the presence of structure and forming planets.
Our results of refractory depletion and timescales for depletion are qualitatively consistent with expectations of dust growth and radial drift including partitioning of elements and constitute direct evidence that radial drift of solids locked in pebbles takes place in disks.
\end{abstract}


\section{Introduction} \label{sec:intro}

Protoplanetary disks are natural outcomes of the star formation process. During their lifetime, the  initially gas-rich disks evolve by accreting mass onto the star and losing it through winds and photoevaporation \citep[e.g.][]{hartmann_accretion_2016}.
Concurrently, the mass locked in solids (dust) grows, settles, and drifts radially to the star, evolving into planetesimals and planets, terrestrial or gas/ice giants \citep{weidenschilling1997,brauer_coagulation_2008,birnstiel2012}.
As a consequence, the distribution of materials and the chemical composition in the disks 
differs from the ones of the parent cloud \citep[e.g.][]{oberg_protoplanetary_2023}.

The abundance of rocky (refractory) material, especially in the innermost gas disk, provides a crucial record of the processes that make the chemical composition change.
Because refractories require the highest temperatures to sublimate, they remain in pebbles -- mm/cm-size 
solid particles --, and are only affected by radial transport or trapping mechanisms in the disk \citep[e.g., radial drift, pressure bumps,][]{drazkowska_planet_2022}. 
One of these processes is
planet formation,
which not only removes the rocky material from the disk as they build their cores \citep[e.g.,][]{lambrechts_rapid_2012,drazkowska_planet_2022}, but also blocks the inward drifting pebbles in pressure bumps \citep[e.g.][]{pinilla_trapping_2012,zhu_dust_2012,vandermarel_diversity_2021}, reducing the abundance of rocky material in the inner regions. 

Metal depletion in young stars has been detected before; in particular, TW Hya studies in the ultraviolet reveled depletion of Fe and Si \citep{Kastner_evidence_2002, Herczeg_faruv_2002, Stelzer_x-ray_2004}. In recent years, the idea that planet-induced disk pressure traps can cause the depletion of refractory elements in the accreting gas and the stars themselves has been explored 
both observationally, by looking at the stellar abundances of Herbig Ae/Be stars \citep[][]{kama_fingerprints_2015, guzman-diaz_relation_2023} and at the composition of gas accreting onto T Tauri Stars \citep[TTSs;][]{mcclure_carbon_2019,mcclure_measuring_2020, micolta_ca_2023} and Brown Dwarfs \citep{france_metal_2010}, and theoretically by exploring the effects of giant-planet formation in disk evolution models \citep[][]{schneider_how_2021a,schneider_how_2021b,huhn_how_2023}.

In an effort to understand the behavior of refractory material in the inner disk and as a tracer of the disk structures, we adopted Ca as a proxy for this material and used the Ca II lines to measure the abundance of Ca to H. In accreting TTSs, the CaII lines are the strongest emission lines of refractory elements present in the optical range and formed in the magnetospheric accretion flows – as evidenced by their higher line luminosities compared to non-accreting stars, and profiles with high velocity wings and red-shifted absorption \citep[e.g][]{azevedo_calcium_2006, Motooka_measurements_2013, micolta_ca_2023} – which feed directly from the inner disks \citep{hartmann_accretion_2016}.
In \citet{micolta_ca_2023}, a systematic analysis of the \CaII\ lines for TTSs of the Chamaeleon I cloud was performed, finding that the \CaII\ lines are much more sensitive to the accretion rate than the \halpha\ line; the \CaII\ line profiles of the lowest accretors resemble the narrow profiles of non-accreting stars, with the chromospheric contribution dominating the emission. The profile analysis also revealed the first hint towards the possibility of underabundance of Ca in the inner gas disks of TTS, as some stars 
showed narrower \CaII\ profiles than expected given their broad and strong \halpha\ lines, high values of the mass accretion rate, and predictions of 
magnetospheric accretion models.

Using an approximate indicator of the abundance of Ca based on the line luminosities, \citet{micolta_ca_2023} found Ca depletion in all disks with cavities in
their spectral energy distributions (SEDs), which are possibly opened by planets \citep{espaillat_observational_2014}, consistent with the results of \citet{Motooka_measurements_2013}, which found that the equivalent widths of the \CaII\ infrared triplet lines in TTSs with gaps were consistently one-tenth smaller than those without cavities.
Furthermore, Ca depletion was found in some stars without signs of gaps in their SEDs. In all cases,  we found that the level of depletion was too high to be explained by differences in the initial conditions. This suggests that Ca depletion in the inner disk could be a tracer for dust-trapping mechanisms that had yet to or would not open large enough gaps for them to show up in the SEDs.  However, this analysis was performed for one star-forming region and the actual Ca abundance relative to H was calculated for only one star, T28. To obtain a broader view of this phenomenon and its possible origins, we need estimates of refractory abundances in the inner disk over a range of masses, ages, and environments.

Here we report the Ca abundances for the inner disks of accreting TTSs in the regions of Chamaeleon I, Lupus, and the Orion OB1b association. We search for correlations between the Ca abundance of the inner disk and the properties of stars and disks. By quantitatively analyzing the {\CaII} lines formed in the accretion flows, we can access the bulk of the refractory material in the inner disk and by comparing our estimates with our sample properties, we aim to gain some insight into the processes responsible for the trapping of refractory elements.

This paper is organized as follows. In Section \ref{sec:Obs} we describe the observations and data sources. In Section \ref{sec:mod} we describe the magnetospheric accretion models.
In Section \ref{methods} we describe the methods used for the analysis of the observations and the application of the models to derive the Ca abundances.
In Section \ref{sec:results} we present our results and explore the correlations with stellar and disk parameters. 
In Section \ref{sec:dis}, we discuss the implications of our results. Finally, in Section \ref{sec:con} we give our conclusions.

\section{Observational Material} \label{sec:Obs}

\subsection{Classical T Tauri Stars} \label{subsec:ctts}

Our sample of accreting TTSs (Classical TTSs, CTTSs) consists of a subset of stars with flux-calibrated spectra from the X-shooter spectrograph \citep[][]{vernet_x-shooter_2011} at the ESO Very Large Telescope (VLT) for three different star-forming regions: Lupus \citep[Lup, ][]{alcala_x-shooter_2014,alcala_x-shooter_2017}, Chamaeleon I \citep[ChaI, ][]{manara_x-shooter_2016,manara_x-shooter_2017}, and the Orion OB1b subassociation \citep[Ori 1b, PENELLOPE VLT, Large Program,][]{manara_penellope_2021}.
For ChaI and Lup we adopted the stellar and accretion parameters of \citet[][]{manara_extensive_2017} and \citet{alcala_x-shooter_2017}, respectively, derived following the methods of \citet{manara_accurate_2013}, using the average distances for each cloud or subcloud. For Ori 1b we adopted the stellar parameters from \citet{manara_penellope_2021}, similarly carried out with the method of \citet{manara_accurate_2013}, but using the GAIA EDR3 \citep{gaia_edr3_2021} distances.

For the accretion properties of the Ori 1b targets, we used the revised values of \citet{pittman_towards_2022} obtained from modeling NUV and optical HST spectra, with the exception of CVSO109 and CVSO165; these are binary targets resolved by HST \citep{proffitt_close_2021} but unresolved in the X-Shooter data; therefore, we used the parameters derived for the combined spectra in \citet{manara_penellope_2021}. 
For each star, we adopt the same distance as the one used to derive the stellar and accretion parameters.

We focus on stars with \halpha\ luminosities above the maximum luminosity observed for WTTS in our range of spectral types, that is $\log\ \left( {L_{H\alpha}}/{\lsun} \right)\ \geq  -3.75$. With this criterion, we ensure that the emission is dominated by accretion and not the chromosphere, for which the analysis based on fluxes is applicable \citep[c.f.,][]{micolta_ca_2023}. 
For the ChaI and Lup targets, we adopt the disk properties compiled in \cite{manara_demographics_2023}, where the disk dust masses were obtained from the ALMA Band 6 or 7 continuum data \citep{pascucci_steeper_2016, ansdell_alma_2016, ansdell_alma_2018, long_alma-chai_2018}, and the disk dust sizes were taken from \cite{Hendler_evolution_2020}. None of the Orion OB1b targets has millimeter imaging observations.
Table \ref{tab:regions} shows the properties of each region, and Table \ref{tab:ctts} shows the properties of each star included in our sample.

Table \ref{tab:ctts} includes information on the known disk morphology of the targets, grouped into three main categories: Full Disks (FD, disks without known cavities or substructures), disks with substructures (SubS, disks with identified substructures by millimeter 
imaging), and Transitional Disks (TD, disk with inner cavities inferred from SED modeling and confirmed by millimeter imaging). We note that most of our targets lack high spatial resolution observations.
\citet{pascucci_steeper_2016} surveyed the ChaI region at a resolution of 0.6", resolving only cavities $\geq$50 au. Similarly, most of the Lupus targets were imaged at 0.25" and only the cavities $\geq$ 20 au have been resolved \citep{ansdell_alma_2016}. 
Therefore, one cannot rule out the presence of smaller gaps or substructures in full disks; for instance, CR Cha appeared as a full disk in \citet{pascucci_steeper_2016}, but an outer ring was identified in \citet{kim_detection_2020}.

\begin{deluxetable}{lcccc}[t!]
\tablecaption{Star-forming regions}
\label{tab:regions}
\tablehead{
\colhead{Region} & \colhead{ d (pc)} & \colhead{Age(Myr)} & \colhead{N} & \colhead{Ref} 
}
\startdata
         Lup   &  150-200 &  3$\pm$2  & 25  & 1,2,3 \\
         ChaI & 160  &  3-5  & 37  & 4,5 \\
         Ori 1b  &  400 & 5 & 8   & 6 \\
          \hline
    \enddata
\tablecomments{ For each region, we report the average age found in literature, except for Lupus where we report the median age. }
\tablerefs{(1) \cite{alcala_x-shooter_2014}, (2)\cite{alcala_x-shooter_2017} (3) \cite{frasca_x-shooter_2017}, (4) \cite{Luhman_Stellar_2007} (5) \cite{manara_x-shooter_2017}, (6) \cite{Briceno_cida_2019}. }
\end{deluxetable}

\subsection{Weak-line T Tauri Stars} \label{subsec:wtts}

Given how magnetically active TTSs are, 
we need to consider the chromospheric contribution to properly compare the results of the magnetospheric accretion models with observations of CTTSs. 
For this purpose, we used X-shooter observations \citep{manara_x-shooter_2013,manara_extensive_2017} of non-accreting TTSs (weak-line T Tauri stars, WTTS) in the same range of spectral types as our CTTS sample. The stellar properties are adopted from \citet{manara_x-shooter_2013,manara_extensive_2017} and the line fluxes from \citet{micolta_ca_2023} (see Table \ref{tab:wtts}).

\section{Model} \label{sec:mod}

\subsection{Magnetospheric Accretion Model} \label{sec:magneto}

We use {\it CV-multi} \citep{muzerolle_emission-line_2001} to calculate the structure and emission of the magnetospheric accretion flows
\citep{muzerolle_emission-line_1998,muzerolle_emission-line_2001,hartmann_magnetospheric_1994}. This framework
assumes that the accretion flows follow the stellar magnetic field lines, which have a dipole geometry characterized by the inner radius (\ri) and the width at the base of the flow (\Dr) in the disk. 
The density distribution is determined by the geometry and the mass accretion rate (\mdot). The models employ a semi-empirical temperature distribution obtained balancing a volumetric heating rate ($r^{-3}$) and an optically thin cooling law \citep[][]{hartmann_magnetospheric_1994}.
It is parametrized by the maximum flow
temperature (\tmax), constrained by the accretion rate, following the prescription of \citet[][]{muzerolle_emission-line_2001}.
The models use the extended Sobolev approximation to calculate mean intensities, which in turn are used to calculate the radiative rates in the statistical equilibrium equations. We adopt a 16-level hydrogen atom \citep[][]{muzerolle_emission-line_2001} and a 5-level calcium atom \citep[][]{micolta_ca_2023} to calculate populations and optical depths. The abundance of Ca relative to H 
in the flows is left as a free parameter,
and we cite it relative to the solar value \citep[$\rm log \left(N_{Ca} / N_{H}\right)_{\odot} = 6.31 \pm 0.04$,][]{asplund_cosmic_2005}.
Line profile calculations are performed using the ray-by-ray method for a given inclination angle $i$ and assuming Voigt profiles \citep{muzerolle_emission-line_2001}.

We calculated a large grid of magnetospheric models for five stars with spectral types covering the range of our CTTS sample. The stellar parameters were taken from the 3 Myr isochrone of the PARSEC evolutionary models \citep[][Table \ref{tab:model_stellar_param}]{bressan_parsec_2012}, consistent with the average ages of our sample (cf. Table \ref{tab:regions}). For each model, we calculated profiles for the Balmer and Ca II lines, obtaining a total of 734,400 profiles combined for all the lines and stars. The full parameter space explored is described in Table \ref{tab:model_param}, where for each combination of stellar and magnetospheric parameters, we calculated
line profiles for abundances (by number) 
$\rm \left(N_{Ca} / N_{H}\right) /\left(N_{Ca} / N_{H}\right)_{\odot} = 1, 0.5, 0.1, 0.01$
or equivalently
[Ca/H]\footnote{$\rm [Ca/H] = log (N_{Ca}/N_{H}) - log (N_{Ca}/N_{H})_{\sun}$, where N refer to the abundances by number.} = 0, -0.3, -1, -2.

\begin{deluxetable}{lccc}[t!]
\tablecaption{Magnethospheric parameter space of the models \label{tab:model_param}}
\tablehead{
\colhead{Parameters} & \colhead{Min.} & \colhead{Max.} & \colhead{Step} 
}
\startdata
SpT	                        & M5	& K2 	& 2$^*$  \\
log {\mdot} ($\msunyr$)	    & -10.0	& -7.0 	& 0.25  \\
T$_{\rm max}$ (K)	        & 6500	& 14000	& 500 \\
R$_{\rm i}$	(R$_{\star}$)	& 2.0	& 6.0	& 0.5 \\
$\Delta\rm r$ (R$_{\star}$)	& 0.5   & 2.0   & 0.5 \\
$i$	(deg)				    & 15	& 75	& 15 \\ 
$[$Ca/H$]$	& -2	&  0	& ... \\ 
\hline
\enddata
\tablenotetext{*}{The last step made in SpT was 3 subclasses to better match the observations range.}
\end{deluxetable}

\begin{deluxetable}{lccc}[t!]
\tablecaption{Stellar parameter space of the models \label{tab:model_stellar_param}}
\tablehead{
\colhead{SpT} & \colhead{Teff (K)} & \colhead{R} & \colhead{M} 
}
\startdata
K2  & 4900   & 1.92      & 1.43     \\
K5  & 4350   & 1.52      & 0.87     \\
K7  & 4060   & 1.45      & 0.73     \\
M1  & 3705   & 1.43      & 0.61     \\
M3  & 3415   & 1.33      & 0.47     \\
M5  & 3125   & 1.17      & 0.31     \\
\hline
\enddata
\end{deluxetable}

\subsection{Model's Grid Reduction} \label{subsec:grid}

After creating the grid of models, we found combinations of parameters do not reproduce line profiles comparable to observations, i.e. the parameters generate line profiles with Balmer lines fully in absorption or dominated by red-shifted absorption,  which do not correspond to the definition of the T Tauri class \citep{Herbig_second_1972}. To properly identify all the combinations of parameters that result in this type of profile across the extensive grid, we evaluated the shape of the profiles by comparing the flux above and below the continuum for each line, obtaining that models reproduce realistic line profiles if
$\rm log(F_{H\alpha}) > 6.5$ and $\rm log(F_{H\beta}) > 6.5$, where $F$ is the line flux above the continuum in $\rm erg s^{-1} cm^{1} Hz^{-1}$ at the stellar surface, and we use this as our selection criterion to find the final models used when comparing with observations.

\section{Methods}\label{methods}

To estimate the abundance of Ca in our targets, we focused on two lines: \halpha\ and \CaIIk. The \halpha\ line traces the bulk of the gas, probing the accretion rate of the star \citep[e.g][]{hartmann_accretion_2016}. The \CaIIk\ line is the strongest of the \CaII\ lines, without contamination from nearby hydrogen lines or significant photospheric contribution,
so it can probe the depletion of Ca with fewer uncertainties \citep{micolta_ca_2023}.

\subsection{Line Fluxes and Luminosities} \label{subsec:lum}

For Cha I stars and WTTS, we used the fluxes reported in \cite{micolta_ca_2023}.
For the Lupus and OB1b stars, we followed the same methods and calculated the fluxes of the \halpha\ and \CaIIk\ lines by integrating the continuum-subtracted line profiles corrected for extinction. The contribution of the photosphere for early K stars was removed by subtracting synthetic BT-Settl spectra \citep[][]{baraffe_new_2015,allard_atmospheres_2012}. For a given star, we used a
BT-Settl model with \teff\ within 50 K of the star and a typical value of log $g = 4.0$. Before subtraction, the model was convolved to the X-shooter resolution and rotationally broadened at the same rotational velocity ($v\sin i$) as the object. We used
$v\sin i$ values from \cite{frasca_x-shooter_2017} for Lupus and \cite{pittman_towards_2022} for Orion OB1b.

To determine the line fluxes, we calculated three independent measurements per line corresponding to the lowest, highest, and middle positions of the local continuum depending on the local noise level of the spectra. Subsequently, the flux and its error were calculated as the average and standard deviation of the three independent measurements, respectively.
All fluxes and their errors are provided in Tables \ref{tab:ctts} and \ref{tab:wtts} for CTTS and WTTS, respectively. The luminosity of the emission lines follows as $L_{\rm line} = 4\pi d^2 \ F_{\rm line}$ where $d$ is the distance to the star (cf. Table \ref{tab:ctts}).

To include the contribution of the chromosphere to the line luminosities of the models, we added the line luminosities of the WTTS of the same spectral type
to those of the models.

\begin{figure*}[t!]
\epsscale{0.92}
\plotone{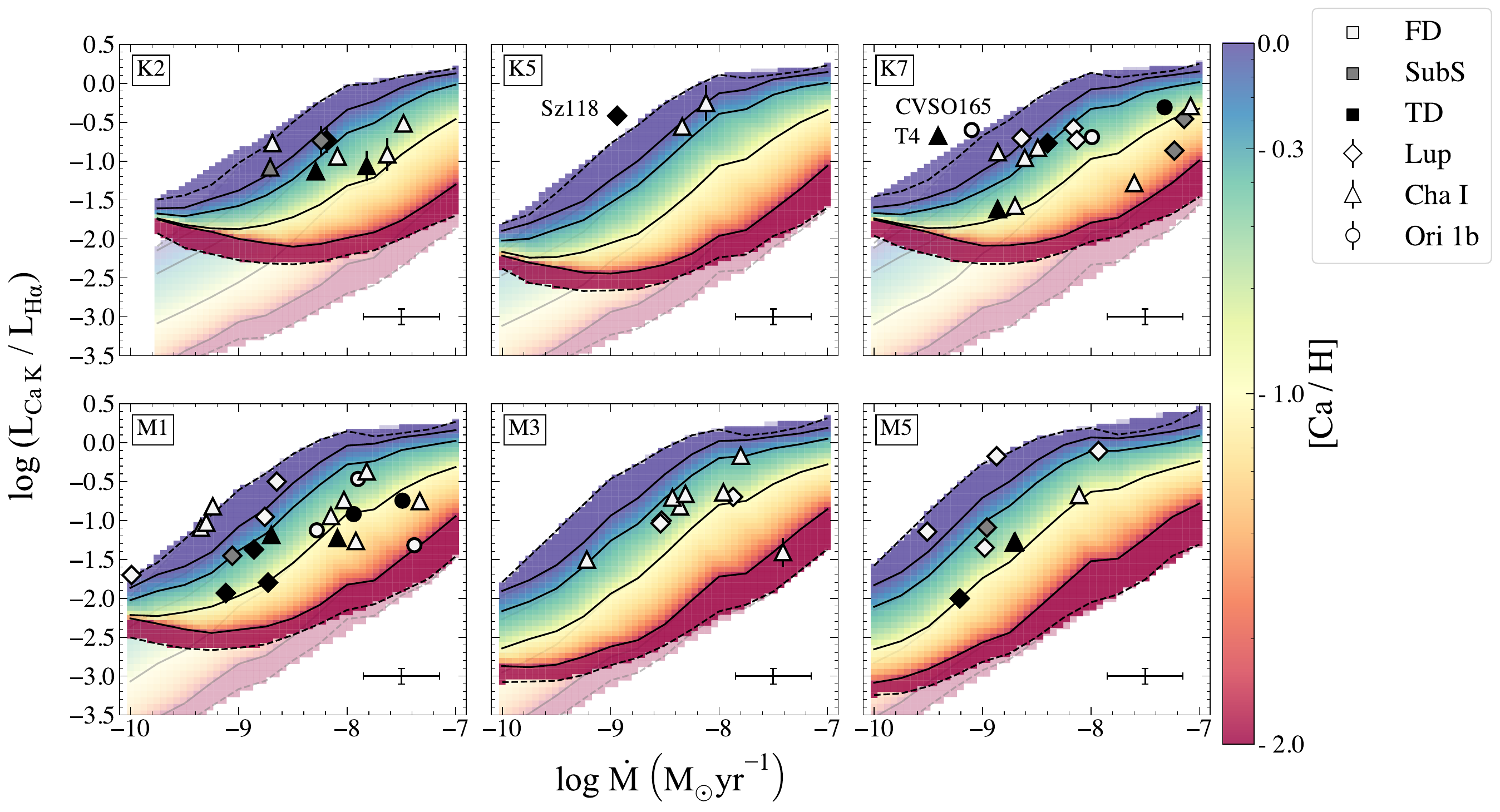}
\caption{Comparison of $\rm log (L_{CaII K} / L_{H\alpha})$ vs. accretion rate between magnetospheric accretion models (background) and observations (dots). Models are colored by $[\rm{Ca/H}]$, with purple representing solar abundance (1) and depletion increasing toward red. The gray-out region represents the magnetospheric models without the contribution of the WTTS. Solid lines represent the median value for a given model abundance (1, 0.5, 0.1, 0.01) and dashed lines represent the 16\% and 84\% percentiles for the lowest and highest abundance values of the models, respectively. Typical uncertainties for observations are shown in the lower right corner.}
\label{fig: grid}
\end{figure*} 

\subsection{$\rm Ca$ abundances} \label{subsec:ca-abund}

To estimate the abundance of Ca, we compared the measured $\rm{L_{Ca\ II\ K}/L_{H\alpha}}$ values with those that would be expected from magnetospheric accretion models at the mass accretion rate {\mdot} of each source. The mass accretion rate traces the density of the accretion flows, that is, the amount of material that falls onto the star, and
it is known to be well correlated with $L_{H\alpha}$ \citep{ingleby_accretion_2013, alcala_x-shooter_2014,micolta_ca_2023}. The \mdot\ values for our targets were measured independently of this work from the UV excess \citep{alcala_x-shooter_2017, manara_extensive_2017, manara_penellope_2021, pittman_towards_2022}. For a star depleted in Ca, 
$\rm{L_{Ca\ II\ K}}$ will be less than expected for its accretion rate \citep{micolta_ca_2023}; therefore, the $\rm{L_{Ca\ II\ K}/L_{H\alpha}}$ ratio will also be smaller than expected for its \mdot.
In the following paragraphs, we describe how we calculated the Ca abundance for our targets.  

We calculated the line luminosities of the models using the fluxes and radii of the model star (Table \ref{tab:model_stellar_param}).
For each of the models, we added line luminosities of a WTTS of the same spectral type to include the chromospheric level. 
Then, we took all the models with the same Ca abundance and mass accretion rate and calculated the median, 16th, and 84th quartiles of the $\rm{L_{Ca\ II\ K}/L_{H\alpha}}$.
This characterizes the behavior of $\rm{L_{Ca\ II\ K}/L_{H\alpha}}$ with respect to the accretion rate for each value of $[\rm{Ca/H}]$. 

To obtain a continuous behavior along the models, we interpolated between the median of the line ratio for each abundance of Ca, including the quartiles for the edges, generating a 3D piecewise interpolant between the $\rm{L_{Ca\ II\ K}/L_{H\alpha}}$  ratio, the accretion rate, and the $[\rm{Ca/H}]$ values of the models. The color grid in Figure \ref{fig: grid} shows the overall resulting behavior of the Ca abundance in the $\rm{L_{Ca\ II\ K}/L_{H\alpha}}$ ratio versus $\rm log\ \Mdot$ diagram; purple represents solar values,  with depletion increasing towards red. Solid black lines represent the median value for each abundance; the dashed lines are the 16th and 84th quartiles for the lowest and highest abundance values, respectively. For comparison, we applied the same steps to the models without the chromospheric contribution, which is shown as the gray-out region in Figure \ref{fig: grid}.

In addition to the models, we show the measured $\rm{L_{Ca\ II\ K}/L_{H\alpha}}$ values for our sample with respect to $\rm log\ \Mdot$. The color of each point indicates whether the disk is known to have gaps (TD, black), a substructure (SubS, gray), or whether it is smooth (FD, white), and the shape represents the star-forming region to which each point belongs.
To estimate the Ca abundance for our observations, we applied the constructed interpolants for the median, 16th, and 84 percentiles of the models, using the measured line ratio and the independently obtained accretion rate. We adopted the average value of the three determinations as the Ca abundance and the standard deviation as the error resulting from the method. Given the coarse nature of this technique, the uncertainties are mainly driven by the method rather than by the intrinsic uncertainties of the stellar parameters. Only the intrinsic error of the accretion rate is non-negligible, we include this uncertainty by estimating the standard deviation of three values, the mean Ca abundance, and the abundance if the star had the accretion rate plus or minus its uncertainty. The general uncertainty is calculated by adding in quadrature; final values and errors are reported in Table \ref{tab:results}.
We have three outliers, namely Sz118, T4, and CVSO165, we discuss the possible reasons for this behavior in \S\ref{subsec:outliers}.

We emphasize that this is a coarse method for calculating abundances; however, it allows us to analyze a large sample in an efficient way. A refined estimate of the abundance values requires simultaneous detailed fitting of multiple line profiles; the general abundance estimates are anticipated to remain the same, while the uncertainties of each estimate will be minimized. This will be carried out in future work for targets of interest.

\section{Results} \label{sec:results}

\begin{figure}[!t]
\epsscale{1.1}
\plotone{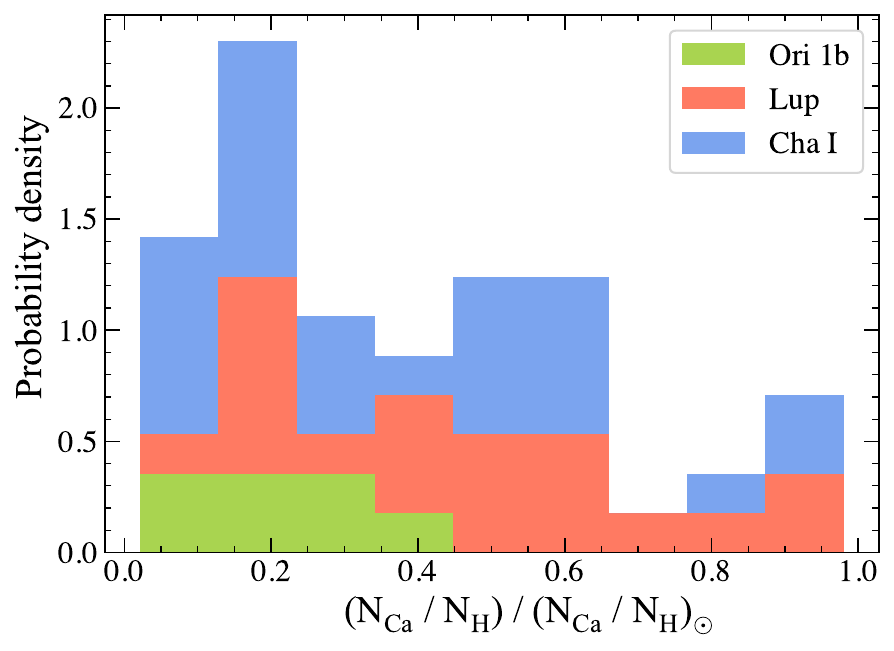}
\caption{Probability density of the abundance by number, colored by star-forming region. }
\label{fig: hist}
\end{figure}

Figure \ref{fig: hist} shows the
probability density of Ca
abundance in Ori OB1b, Lupus, and Chamaeleon. 
In general, we find a wide range of Ca abundances in all three star-forming regions.
The overall distribution is right-skewed, with 57\% of the sample having abundances $\rm \left(N_{Ca} / N_{H}\right) /\left(N_{Ca} / N_{H}\right)_{\odot} < 0.50$ or $ [\rm{Ca/H}] = -0.30$, relative to solar, suggesting that there is a tendency toward high Ca depletion values (low abundances) in CTTS. 

We explored the probability that the distributions of the three star-forming regions were sampled from populations with identical distributions by performing two-sample Kolmogorov-Smirnov (KS) tests, where the null hypothesis was that the two distributions were identical, F(x)=G(x) for all x. The p-value gives the probability of obtaining the observed results assuming that the null hypothesis is true. For the purposes of this work, we rejected the null hypothesis if the value $p$ of the chosen statistic was $<$ 0.05 ($5\%$).

We performed three KS tests, using the full sample of each region: ChaI-Lup, ChaI-OB1b and Lup-OB1b (Table \ref{regionsstats}). For the
first two cases, we obtain $p$-values significantly above 5\%, indicating that it is unlikely that the underlying abundance distributions of ChaI-Lup and ChaI-Ob1b differ from each other. The $p$-value for the Lup-OB1b test suggests that the probability for the underlying distributions of Lup and OB1b are the same is  $\sim$10\% . Except for CVSO165, which falls outside the parameter space of the models (see \S \ref{subsec:outliers}), all stars in our OB1b sample are depleted, with a maximum value of $\rm \left(N_{Ca} / N_{H}\right) /\left(N_{Ca} / N_{H}\right)_{\odot}$ of 0.41 or $ [\rm{Ca/H}] = -0.39$. As OB1b is the oldest region and Lupus the youngest of the sample, this could indicate age-dependent depletion. However, we cannot reach a strong conclusion,
since the null hypothesis cannot be rejected, and also considering that the Ori 1b sample is small compared to those of ChaI and Lup. Therefore, we cannot rule out the possibility that the underlying distributions of the three regions coincide with each other. Larger samples of old populations are needed to provide a more definitive conclusion.

\begin{deluxetable}{lccclll}[t!] \label{regionsstats}
\tablecaption{ $\rm \left(N_{Ca} / N_{H}\right) /\left(N_{Ca} / N_{H}\right)_{\odot}$ by regions KS test}\label{tab:mdust-stats}
\tablehead{
\colhead{ - } &\colhead{$statistic$} & \colhead{$p$-value}
}
\startdata
ChaI~-~Lup  & 0.23 & 0.47 \\
ChaI~-~Ori 1b & 0.34 & 0.40 \\
Lup~-~Ori 1b & 0.48 & 0.10 \\
\enddata
\end{deluxetable}

\subsection{Exploring relationships between Ca abundance and stellar and disk parameters} \label{subsec:corr}

In the next subsections, we explore correlations between the obtained Ca abundances and stellar parameters and disk structure for the whole sample. Additionally, we explore possible relationships between the abundance and and disk properties for the ChaI and Lupus sub-samples, for which disk parameters are known.  We discuss the most significant insights in Section \ref{sec:dis}.

\subsubsection{Relationship with stellar Mass}\label{sec:mass}

\begin{figure}[!t]
\epsscale{1.1}
\plotone{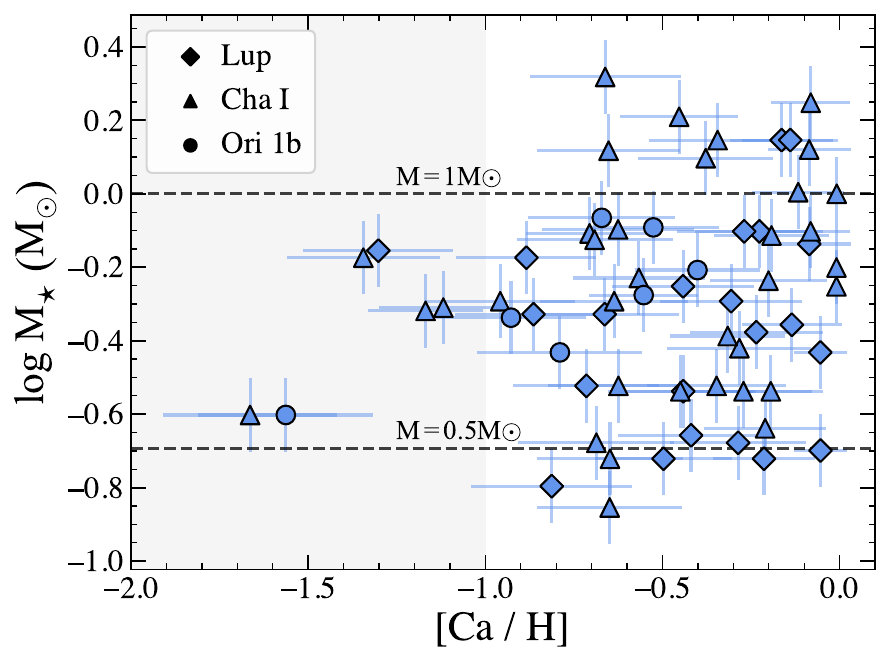}
\caption{$[\rm{Ca/H}]$ vs. $\rm M_{\star}$  The grey region highlights the$[\rm{Ca/H}] \leq -1$ region, the minimum abundance we are able to detect for $\rm M_{\star} > 1$. No relationship between the stellar mass and Ca abundance is found.}
\label{fig: stellarmass}
\end{figure}

Figure \ref{fig: stellarmass} shows the abundance of Ca ($[\rm{Ca/H}]$) versus the stellar mass ($\rm M_{\star}$). We note that for the $\rm M_{\star} > 1 M_{\odot}$ sample, we do not find abundance values lower than $-1$ (gray area). At first glance, this would suggest that the lowest abundance values are found only in low-mass stars. However, biases in our estimates due to the small size of the high-mass sample and the detection limit imposed by the chromosphere probably contribute to the observed gap in abundances (see \S\ref{subsec:bias}, cf. Fig. \ref{fig: grid}).

We explored the statistical significance of the difference observed within the two mass groups ($\rm M_{\star} \leq 1 M_{\odot}$ and $\rm M_{\star} > 1 M_{\odot}$), using a two-sample KS test, considering stars with $[\rm{Ca/H}] > -1$, to mitigate biases. We obtained a statistic of 0.33 with a $p$-value of 0.26, which provided evidence for the null hypothesis, that is, that the underlying distributions of the two mass samples were the same.

To test for a correlation between abundance and stellar mass, we calculate the Pearson correlation coefficient (hereafter the Pearson $r$ test). In the cases where the $p$-value is $<0.05$, we reject the null hypothesis that there is no statistically significant relationship between the variables. We obtain a statistics of 0.17 with a $p$-value of 0.20; therefore, we do not find a correlation between Ca abundance and stellar mass.

\begin{deluxetable*}{lcccccc}[t!]
\tablecaption{$[\rm{Ca/H}]$ vs. Properties Statistical Tests}\label{tab:stats}
\tablehead{
\colhead{} & \multicolumn{2}{c}{Pearson r} & \multicolumn{4}{c}{Regresion Parameters}\\
\cmidrule(lr){2-3} \cmidrule(lr){4-7}
\colhead{ $[\rm{Ca/H}]$ vs. } &\colhead{$r$} & \colhead{$p$-value} & \colhead{$\alpha$} & \colhead{$\beta$}& \colhead{$\sigma$} & \colhead{$\hat{\rho}$} 
}
\startdata
log $\rm M_{\star}$  & 0.17 & 0.20 & ... & ... & ... & ... \\
log \mdot  & -0.58 & 5.46E-07 & 
-8.86$^{+0.12}_{-0.12}$ & -1.30$^{+0.24}_{-0.24} $& 0.07$^{+0.05}_{-0.05}$ & -0.86$^{+0.10}_{-0.10}$ \\
log $\rm M_{dust}$  & -0.16 & 0.23 & ... & ... & ... & ... \\
log $\rm R_{68}$  &  -0.17  & 0.33  & ... & ... & ... & ...\\
\enddata
\tablecomments{The regression parameters fit $Y(X) = \alpha + \beta\ X$, where $\alpha$ and $\beta$ are the intercept and slope, respectively.  $(\sigma)$ is the scatter of the relation $(\sigma)$ and $(\hat{\rho})$ the correlation coefficient.}
\end{deluxetable*}

\begin{figure}[t!]
\epsscale{1.1}
\plotone{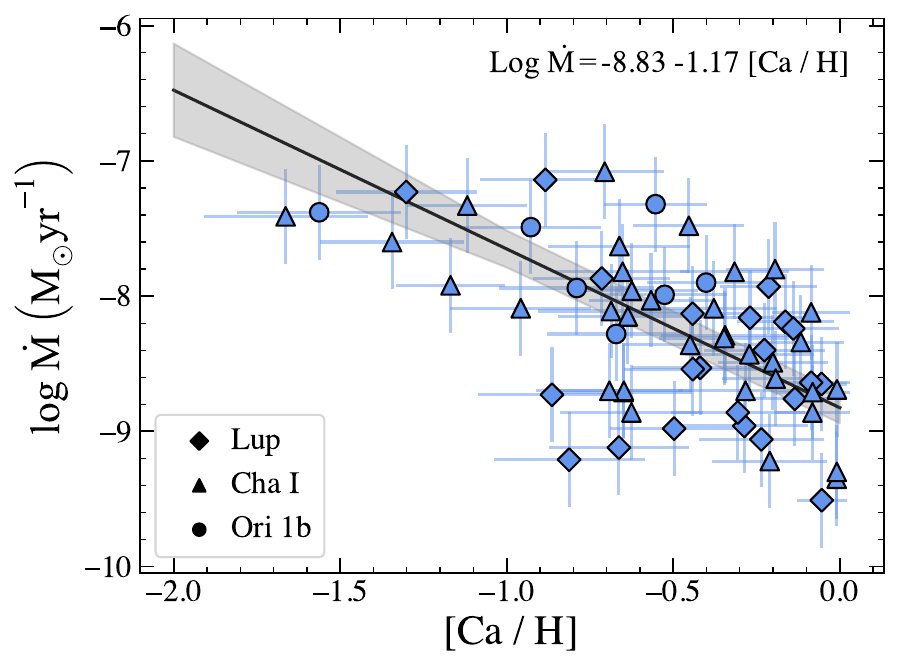}
\caption{$[\rm{Ca/H}]$ vs Log(\mdot). The black line represents the best fit from the MCMC linear regression; our $\rm 1\sigma$ confidence intervals are shown in gray.}
\label{fig: CavsMdot}
\end{figure}

\subsubsection{Relationship with Mass Accretion Rate}

Figure \ref{fig: CavsMdot} shows the mass accretion rate onto the star (\mdot) versus the Ca abundance for the three star-forming regions considered here. We explore the correlation between the two variables, calculating the Pearson correlation coefficient. We find a $p$-value lower than 5E-6\% (see Table \ref{tab:stats}) and negative statistics ($r$), indicating that the data are anti-correlated, with higher abundances for lower accretion rates. As a follow-up, we then use the Bayesian linear regression method described in \cite{kelly_some_2007} (implemented in the \texttt{ linmix} Python package by Joshua E. Meyers) to fit the linear regression between $[\rm{Ca/H}]$ and log {\mdot} (Fig. \ref{fig: CavsMdot}).
This procedure takes into account the upper limits and intrinsic scatter in the data, fitting for $Y(X) = \alpha + \beta\ X$, where $\alpha$ and $\beta$ are the intercept and slope, respectively. The best-fit parameters, together with the scatter of the relation $(\sigma)$ and the correlation coefficient $(\hat{\rho})$, are reported in Table \ref{tab:stats}. We find $\rm log\ \Mdot \ = -8.83\ -1.18\ [\rm{Ca/H}]$. 
We discuss the implications of this result on \S\ref{sec:dis_mot}

\subsubsection{Relationship with disk structure} \label{sec:struct}

\begin{figure}[!t]
\epsscale{1.1}
\plotone{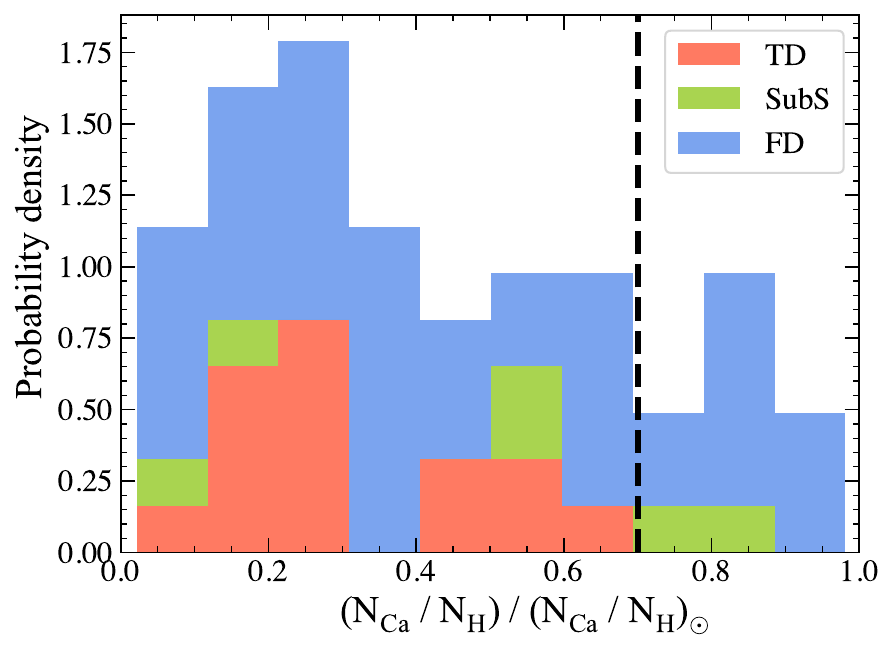}
\caption{Probability density of abundance by number, colored according to the presence of gaps(red, TD), substructures (green, SubS), or absence of both (blue, FD)  in each system. The black dashed lines represents $\rm \left(N_{Ca} / N_{H}\right) /\left(N_{Ca} / N_{H}\right)_{\odot} = 0.7$.}
\label{fig: hist-struc}
\end{figure}

Figure \ref{fig: hist-struc} 
shows the probability density of Ca abundance relative to H (by number), separating disks by their morphology:
presence of gaps inferred from SEDs or imaging (red, TD), substructures inferred from imaging (green, SubS) or non-detection of structure (blue, FD) in each system. 
We explore possible differences between the underlying distributions applying three KS tests, FD-TD, FD-Sub, and TD-Sub. In all cases, we obtain $p$-values significantly above 5\% (see Table \ref{tab:ks-struc}),  indicating the possibility that the null hypothesis cannot be rejected and the underlying Ca abundance distributions coincide with each other. However, this result is affected by the small sample size of transitional and structured disks (17 and 6 stars, respectively) compared to the full disks sample (48 stars). Moreover, the  overall distributions, and therefore this result, are also subject to change as more
cavities or substructures not yet identified in our objects are found. In the next paragraphs we highlight the main takeaways.

The distribution plots show that all known disks with gaps
(TD) for which we obtained an estimate $[\rm{Ca/H}]$ value, show some degree of depletion with $\rm \left(N_{Ca} / N_{H}\right) /\left(N_{Ca} / N_{H}\right)_{\odot} < 0.7$ or $[Ca/H] < -0.15$ (black dashed line in Fig. \ref{fig: hist-struc}), as expected, since cavities act as dust traps on the disk \citep[e.g.][]{pinilla_trapping_2012, zhu_dust_2012, vandermarel_diversity_2021} resulting in lower refractory abundances in the innermost disk \citep[e.g.][]{huhn_how_2023}. Disks with substructures have a wide range of abundances,
indicating that smaller substructures can also significantly impact the amount of refractory material reaching the star. 

The two stars that show substructures in their disks and have $\rm \left(N_{Ca} / N_{H}\right) /\left(N_{Ca} / N_{H}\right)_{\odot} > 0.7$ are Sz68 and CR Cha. Sz68 (HT Lup) is a triple stellar system, with a projected separation of 25 au for the closest pair.
The disk around Sz68 has an underlying spiral structure; however, the spirals are quite compact and appear to connect to the inner disk through a bar-like structure \citep{kurtovic_disk_2018}. CR Cha has a gap at 90 au; inside this gap, the disk does not show evidence of any other substructure. Therefore, a possible explanation for these stars having high abundance values is that even when they have substructures, these are not interrupting the flow of the refractory material drifting towards the star. We note that in our sample Sz 71 (GW Lup), has a substructure similar to CR Cha, a smooth disk with a gap at 74 au \citep{huang_disk_2018} but appears to be depleted; however, for this particular target an unresolved inner cavity or dust trap was proposed to explain the observed column density ratio of CO$_2$ to H$_2$O \citep{grant_minds_2023}; this inner cavity could explain why GW Lup is depleted in Ca but not CR Cha.

What stands out in Figure \ref{fig: hist-struc} is the wide range of $[\rm{Ca/H}]$ values obtained for disks without known gaps/structures (full disks, FD), with abundances ranging from solar to -2. As discussed in \S\ref{subsec:ctts}, we cannot rule out the possibility that some systems may have gaps and substructures not properly identified. This is especially true for the ChaI sample, as only cavities $\gtrsim 50$ au have been resolved \citep[][]{pascucci_steeper_2016} and lack detailed SED modeling. The analysis of Herbig Ae/Be stars made by \citet{kama_fingerprints_2015} identified two stars, HD 142666 and HD 144432, with significant subsolar abundances of refractory elements.
These stars did not show indication of cavities in their SEDs \citep{maaskant_pah_2014}; however, they showed radial gaps on the (sub)au scales in near-infrared interferometry \citep{chen_near_2012,Schegerer_multiwavelength_2013,menu_structure_2015}.
However, we see full disks with high depletion in the Lupus region, one of the better characterized star-forming regions, suggesting that it is not just an issue of low spatial resolution but also a consequence of dust evolution on the disk.

\begin{deluxetable}{lccclll}[t!] \label{tab:ks-struc}
\tablecaption{ $\rm \left(N_{Ca} / N_{H}\right) /\left(N_{Ca} / N_{H}\right)_{\odot}$ by disk structure KS test}
\tablehead{
\colhead{ - } &\colhead{$statistic$} & \colhead{$p$-value}
}
\startdata
FD-TD  & 0.25 & 0.32 \\
FD-Sub & 0.23 & 0.45 \\
TD-Sub & 0.37 & 0.90 \\
\enddata
\end{deluxetable}

\begin{figure}[!t]
\epsscale{1.05}
\plotone{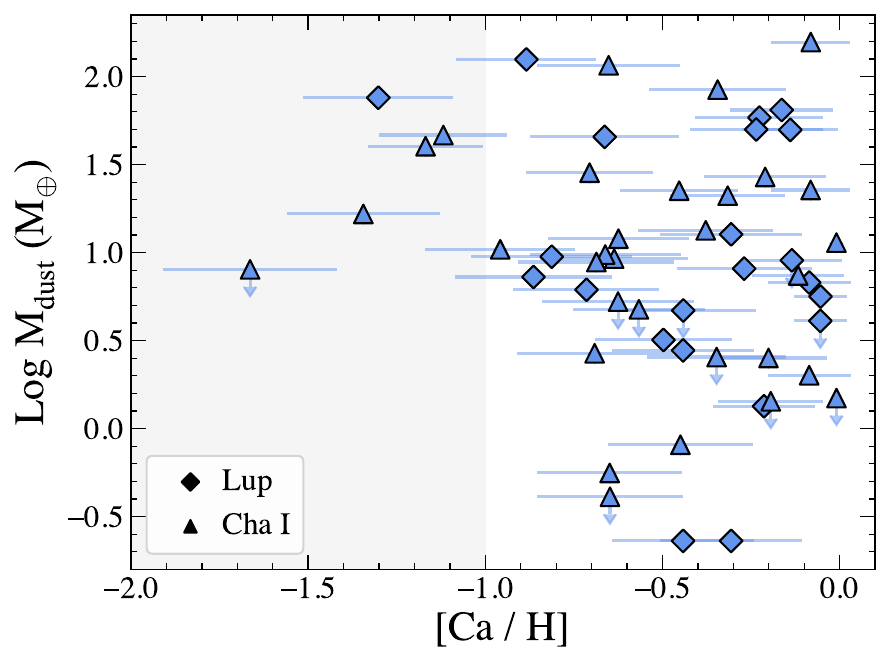}
\caption{$\rm M_{dust}$ vs. $[\rm{Ca/H}]$. Upper limits are indicated as arrows. No relationship between $\rm M_{dust}$ and $[\rm{Ca/H}]$ is found.}
\label{fig: MdvsCa}
\end{figure}

\subsubsection{Relationship with disk dust mass and radius} \label{sec:mdust}

Sub-milliliter observations have been obtained for stars in Cha I and Lupus \citep[e.g.][cf. \S\ref{subsec:ctts}]{ansdell_alma_2016,pascucci_steeper_2016}, from which dust disk masses and radii have been inferred. Figure \ref{fig: MdvsCa} shows
the dust disk mass $\rm M_{dust}$ versus $[\rm{Ca/H}]$ for the ChaI and Lup stars with data from \citet[][and references therein]{manara_demographics_2023}. We apply Pearson's $r$ test and find that there is no statistically significant correlation between $\rm M_{dust}$ and $[\rm{Ca/H}]$ ($p$-value $> 0.05$, see Table \ref{tab:stats}). However, we find that all six stars with $[\rm{Ca/H}] \leq -1$, indicated in Figure \ref{fig: MdvsCa} with the gray region.
-- namely Sz98, T40, T28, VW Cha, T49 -- 
have high $\rm M_{dust}$ values. None of the ChaI stars have predicted or confirmed cavities or substructures. However, Sz98 shows annular substructures in millimeter dust continuum emission imaging but not deficit in the SED, typical of transitional disks \citep[][]{gasman_sz98_2023}. This also supports the notion that smaller substructures can also significantly impact the amount of refractory material reaching the star.

Finding the lowest abundances among the disks with high $\rm M_{dust}$ is consistent with the anticorrelation between Ca abundance and \mdot\ (cf. Fig. \ref{fig: CavsMdot}), since \cite{manara_demographics_2023} find a positive linear relationship between $\rm M_{dust}$ and \mdot\ for young stars of multiple star-forming regions.
Larger disk masses also allow for an easier and faster formation of giant planets \citep[e.g][]{Savvidou_how_2023}, which in turn can lead to the blocking of Ca in the outer disk becoming efficient at an earlier time, when the accretion rate onto the star is still large.
However, we note that some disks with a high dust masses also have a high Ca abundance value, suggesting that there must be an additional parameter that determines
the abundance. 

\begin{figure}[!t]
\epsscale{1.05}
\plotone{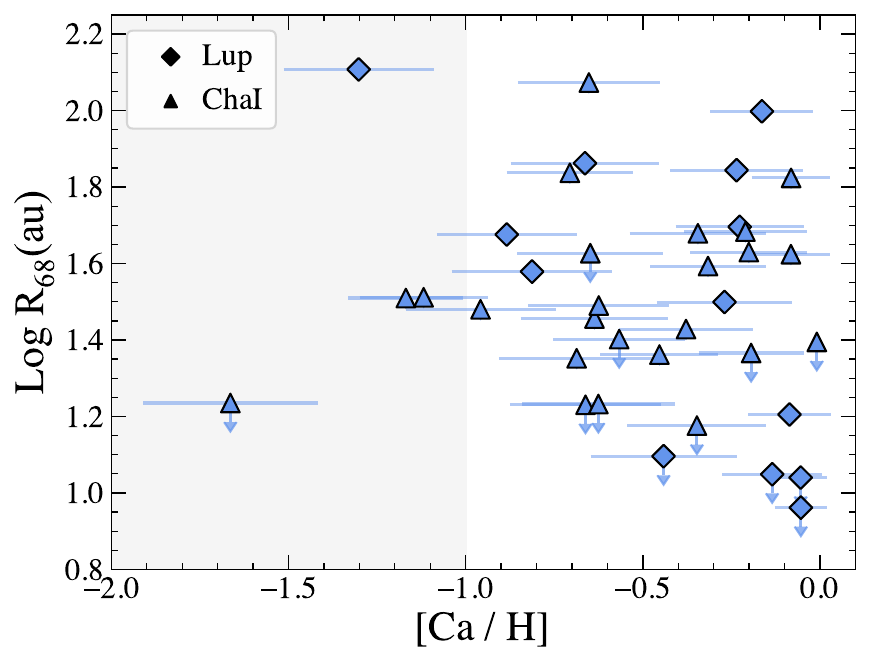}
\caption{$[\rm{Ca/H}]$ vs $\rm R_{68}$. No relationship between $\rm R_{68}$ and $[\rm{Ca/H}]$ is found.
}
\label{fig: R68vsCa}
\end{figure}

Next, we explore a possible correlation of the dust disk size. For this purpose, we use $\rm R_{68}$, the radius containing 68\% of the flux \citep{Hendler_evolution_2020,manara_demographics_2023}. Figure \ref{fig: R68vsCa} shows $[\rm{Ca/H}]$ vs $\rm R_{68}$.
We search for correlation within the variables applying Pearson's $r$ test. We find no correlation between $[\rm{Ca/H}]$ vs $\rm R_{68}$, obtaining a Pearson r statistic of -0.17 with a $p$-value of 0.33 for the whole sample (cf. Table \ref{tab:stats}).

\section{Discussion} \label{sec:dis}

\subsection{Outliers} \label{subsec:outliers}

In Figure \ref{fig: grid} we show that the observed ratios $\rm log (L_{CaII K} / L_{H\alpha})$ when plotted against the mass accretion rates fall within the model predictions if the Ca abundance is allowed to vary.
Three stars fall outside the model space
in Figure \ref{fig: grid}: Sz118, T4, CVSO165. Here, we discuss possible reasons for the higher than expected $\rm log (L_{CaII K} / L_{H\alpha})$ ratio of these stars,
which the models do not reproduce. In principle, models with higher abundance could potentially explain these observations, but overabundances relative to solar are unexpected in this sample because: (1) the star-forming regions are known to have solar abundances within uncertainties \citep[e.g.][]{santos_chemical_2008,spina_gaia-ESo-2014,dorazi_metallicity_2009,biazzo_chemical_2011,biazzo_elemental_2011,Nieva_present_2012}; and (2) the estimated ages for these regions locate them after the initial period of enhanced abundances due to radial drift, predicted in partition models of dust evolution \citep{huhn_how_2023}. Here we discuss other possible reasons why our models do not reproduce these observations.

CVSO165 is a binary system \citep{proffitt_close_2021}, but the X-Shooter spectra 
\citep{manara_penellope_2021}
corresponds to  the combined spectra of two stars, and so we used the stellar parameters derived in \cite{manara_penellope_2021} for the unresolved system. However, 
\cite{pittman_towards_2022} separated both components, and 
derived stellar and accretion properties for both stars. We note that for the more massive component (CVSO165A), the spectral type falls within the uncertainties of the spectral type derived for the unresolved spectra, but the accretion rate is a factor of 10 higher. 
If we assume that the emission line ratio is dominated by the more massive component, then with the higher \mdot\, and the observed $\rm log (L_{CaII K} / L_{H\alpha})$ the star would be located inside the model space, with an abundance value of $\rm \left(N_{Ca} / N_{H}\right) /\left(N_{Ca} / N_{H}\right)_{\odot} \sim  0.43$.

T4 is one of our lowest accretors and falls on the border of our threshold for a significant chromospheric contribution, i.e., when the luminosities of CTTS and WTTS are comparable ($\log\left( {L_{H\alpha}}/{\lsun} \right)\ \geq  -3.75$). Its line profiles also resemble those of a WTTS; therefore, a flux-based analysis is not appropriate for this star. Instead, a detailed profile modeling is necessary \citep[e.g.][]{thanathibodee_censusII_2023}; this will be addressed in future work.

Sz118 is a low accretor that shows multiple components in \halpha: the profile is double-peaked with deep red-shifted absorption that does not go below the continuum at low velocities ($\leq 150\kms$). We attribute this behavior to a highly complex composite magnetosphere \citep[e.g.,][]{thanathibodee_complex_2019}. An accurate modeling of this type of profile is outside the scope of this paper. 

\subsection{Possible sources of bias} \label{subsec:bias}

\subsubsection{Sample sizes}

We note that our sample distribution is skewed towards low-mass stars, that is, the number of high-mass stars is small in comparison to that of low-mass stars. The bulk of our sample is formed by ChaI and Lup stars. Both regions have surveys of Class II objects with completion values of $\sim$90\% for TTSs, where the majority stars have SpT M3-M6
\citep[]{alcala_x-shooter_2017,manara_extensive_2017,micolta_ca_2023},
an outcome of the initial mass function (IMF), which predicts an overdensity of low-mass stars compared to high-mass stars \citep[e.g.][]{bastian_imf_2010}.
Our selection criterion, $\log\ \left( {L_{H\alpha}}/{\lsun} \right)\ \geq  -3.75$
reduces the sample; however, it mainly affects the number of low-mass stars, rather than high-mass stars, as we are effectively excluding the lowest accretors \citep{hartmann_accretion_2016}.

\subsubsection{Detection limits due to the chromosphere}

The chromosphere is an important component of the TTS spectra. In effect, the chromosphere acts like a \textit{floor}, setting the minimum flux that we can detect for a given line, 
and its contribution increases with the effective temperature of the stars. Importantly, the \CaII\ lines are more sensitive to the chromosphere than \halpha; therefore, the increase in \CaIIk\ luminosity is steeper than the increase in the \halpha\ luminosity as we move toward earlier spectral types. We quantified this effect by calculating the Pearson r test and linear regression of the line luminosities versus \teff\ for the entire WTTS sample, following the same steps as in \S\ref{subsec:corr}. The results are shown in Table \ref{tab:wtts-stats}, the slope of $\rm log\ L_{CaII K}$ versus \teff\ is approximately two times the slope of $\rm log\ L_{H\alpha}$ versus \teff. Consequently, the chromospheric contribution in $\rm log (L_{CaII K} / L_{H\alpha})$ becomes important in the earlier spectral types as well, especially low accretion rates as the accretion component decreases. This effect can be seen in Figure \ref{fig: grid}, as the gray-out region increases in area as we move from M5 to K2.

As a result, there is a minimum abundance that we are able to measure over the chromosphere. In particular, for K2 stars and $\rm log\ \Mdot<-8.0$ we are unable to detect abundances $[\rm{Ca/H}]$ lower than $-1$, that is, 
$\rm \left(N_{Ca} / N_{H}\right) /\left(N_{Ca} / N_{H}\right)_{\odot} < 0.1$, using only line fluxes. This, combined with the mass accretion rates of our high-mass sample, can partly explain why we do not see stars with $\rm \left(N_{Ca} / N_{H}\right) /\left(N_{Ca} / N_{H}\right)_{\odot} < 0.20$ for this mass group (see \S\ref{sec:mass} and Figure \ref{fig: stellarmass}).  For spectral types K5 to M1, we cannot estimate abundances lower than $-1$ if $\rm log \Mdot < -8.75$.  For M3 - M5 stars, we should be able to detect abundances lower than -1 for accretion rates $ \geq -9.5$ (cf. Fig. \ref{fig: grid}). 

\begin{deluxetable*}{lcccccc}[t!]
\tablecaption{Line Luminsities vs. effective temperature for WTTS}\label{tab:wtts-stats}
\tablehead{
\colhead{} & \multicolumn{2}{c}{Pearson r} & \multicolumn{4}{c}{Regresion Parameters}\\
\cmidrule(lr){2-3} \cmidrule(lr){4-7}
\colhead{ - } &\colhead{$r$} & \colhead{$p$-value} & \colhead{$\alpha$} & \colhead{$\beta$}& \colhead{$\sigma$} & \colhead{$\hat{\rho}$} 
}
\startdata
$\rm log\ L_{CaII K}$  & 0.88 & 1.31E-09 & $-43.24^{+4.42}_{-4.42}$ & $10.67^{+1.23}_{-1.23} $& $0.91^{+0.05}_{-0.05}$ & $0.16^{+0.07}_{-0.07}$ \\
$\rm log\ L_{H\alpha}$ & 0.75 & 8.44E-06 & $-23.40^{+3.61}_{-3.61}$ & $5.25^{+1.00}_{-1.00} $& $0.77^{+0.10}_{-0.10}$ & $0.13^{+0.05}_{-0.05}$ \\
\enddata
\tablecomments{The regression parameters fit $Y(X) = \alpha + \beta\ X$, where $\alpha$ and $\beta$ are the intercept and slope, respectively.  $(\sigma)$ is the scatter of the relation $(\sigma)$ and $(\hat{\rho})$ the correlation coefficient.}
\end{deluxetable*}

\subsection{Comparison with predictions of refractory depletion}
\label{sed:depletionModels}

We estimated the abundance of refractory elements reaching the star using Ca as a proxy for refractory material. In the disk, Ca is locked in minerals with the highest condensation temperatures \citep{lodders_solar_2003}, so it should remain solid until it reaches the innermost disk, where the temperature is high enough for all material to be in gas. After the disk is truncated by interactions with the stellar magnetic field, gas falls onto the star along the stellar magnetic field lines, emitting characteristically broad spectral lines, which we have used to determine Ca abundances relative to H. 

We have estimated abundances for 70 objects in three different star-forming regions, Cha I, Lupus, and Orion OB1b, covering 1 to 5 Myr, an interesting range of ages in which disk frequencies decrease from $\geq$ 60\% to 10\%
\citep{hernandez_spitzer_2007,ribas_disk_2014,richert_circumstellar_2018}. In addition to a range of ages, our samples cover a range of stellar masses, mass accretion rates, dust disk masses, and disk structures (Table A1).  This diverse sample has allowed us to test for correlations with different parameters in order to explain the origin of the observed depletions.

Currently, we do not have detailed model predictions to compare our observed abundances. \citet[][]{huhn_how_2023} have developed partitioning models of dust evolution in disks, including grain growth, radial drift \citep{birnstiel_dust_2009}, and separation of solids into chemical species. With this treatment, the abundances of elements locked into different minerals change as they drift in and sublimate into the gas phase when they reach temperatures higher than the condensation temperature of the mineral. 

\citet{huhn_how_2023} main emphasis is to make predictions for the abundance of elements on the stellar surface, taking into account the merging of the freshly accreted material with that of the star. In the range of mass and age of our sample, most stars have deep convection zones that are very efficient in mixing the accreted material with the stellar gas, rapidly erasing any chemical signature on the stellar surface \citep{jermyn_stellar_2018, kunitomo_revisiting_2018}. The predictions are applicable to earlier type stars, which lack {\bf sub-photospheric} convective regions, and the accreted material reaching the surface does not get quickly mixed in.

\citet{huhn_how_2023} do show the radial dependence of abundances as a function of age for three elements, C, O and Fe. In their calculation, Fe is locked in minerals with condensation temperature $\sim$ 300 – 700 K, lower than that of the most refractory elements, $\sim$ 1500 -1700K, but still in the inner disk, so its behavior is representative of the refractory elements. The Fe behavior is clearly distinct from that of volatiles C and O. 
In the models shown in Figure 3 of \citet{huhn_how_2023}, the Fe abundance is enhanced relative to solar inside 1 au in the first $\sim$ 2 Myr, but quickly drops below solar beyond 3 Myr. The early overabundance is due to the arrival of the solid material locked in pebbles and moving inward faster than the gas; this material is accreted onto the star, after which the entire disk becomes depleted of Fe. This situation is in sharp contrast to the abundance of the volatiles, which stays above solar even beyond 10 Myr, and only gets depleted in the outer disk. In this case, the fraction of C and O locked in species of low condensation temperature ($\sim$ 20 - 150 K) returns to the gas and takes longer than the refractories to reach the inner disk. 

Our results are consistent with these findings. We find a significant level of depletion in the majority of the stars of the full sample, and, moreover, all the disks in the older sample show depletion (cf. Fig. \ref{fig: hist}, Ori 1b). This suggests that refractory depletion may be a general phenomenon in T Tauri stars, most likely caused by dust evolution. In any event, our results in terms of depletion and timescales for depletion are consistent with expectations of dust growth and drift and constitute direct proof that radial drift of solids locked in pebbles takes place in disks.

The large dispersion in the Ca/H abundance found for each region -- approximately of a given age -- could be due to differences in the initial conditions, the viscosity parameter, and other factors that affect evolution. 
In fact, T Tauri stars in a given population cover a wide range of disk properties, such as mass accretion rates, dust masses, and radii
\citep{hartmann_accretion_2016,rilinger_determining_2023}.
Planet formation may be an additional factor that contributes to the large dispersion in a given population, since the gaps opened by the planets can act as pressure traps \citep{paardekooper_dust_2006,pinilla_trapping_2012,zhu_dust_2012,vandermarel_diversity_2021} and the planets themselves can capture solids
\citep{lambrechts_rapid_2012,thiabaud_stellar_2014,drazkowska_planet_2022}. The fact that all disks with detected gaps and structure, likely due to forming planets \citep[e.g.][]{Zhang_disk_2018}, show depletion supports this statement. 

Partitioning models of dust evolution including refractory elements with high condensation temperature, covering
a wide range of parameters, and allowing for planet formation, are needed for a more detailed comparison with our abundance estimations.  This comparison will provide insight into dust evolution and planet formation models.

\subsection{Dependence of [Ca/H] with mass accretion rate}\label{sec:dis_mot}

Our large sample has allowed us to search for dependences of refractory depletion with stellar, accretion, and disk parameters. We confirm our previous finding \citep{micolta_ca_2023} that refractory depletion is related to the structure of the disk, in the sense that all disks with large cavities identified in the SEDs and images show depletion, although the degree of depletion in structured disks varies. We found no correlation of refractory depletion with stellar mass, disk dust mass, or radius. 

We find a significant correlation of Ca depletion with the mass accretion rate.
This dependence is puzzling, because the mass accretion rate depends on stellar mass as
\mdot\ $\rm \propto M^2$ \citep[][and references therein]{hartmann_accretion_2016} and on dust disk mass \citep[e.g.][]{mulders-constraints_2017, manara_x-shooter_2016,manara_demographics_2023}, and we find that depletion depends on {\mdot}, but does not correlate with either stellar or dust disk mass. We note that in \cite{manara_demographics_2023} the dependencies between stellar and dust disk mass become weak when mixing multiple regions; however, 
we found no 
correlation between {\mdot} and these parameters 
when exploring the regions individually.

The abundance of Ca in relation to H is basically found by comparing the fluxes of the {\CaIIk} line and H$\alpha$. In turn, the flux of \halpha\ is correlated with the mass accretion rate \citep{ingleby_accretion_2013}. Therefore, the correlation between [Ca/H] and {\mdot} with slope $\sim$ - 1 could be understood if the abundance of solids reaching the inner disk was approximately constant, regardless of the abundance of gas reaching the inner disk, assuming that Ca is a proxy for solids. 

\begin{figure}[!t]
\epsscale{1.05}
\plotone{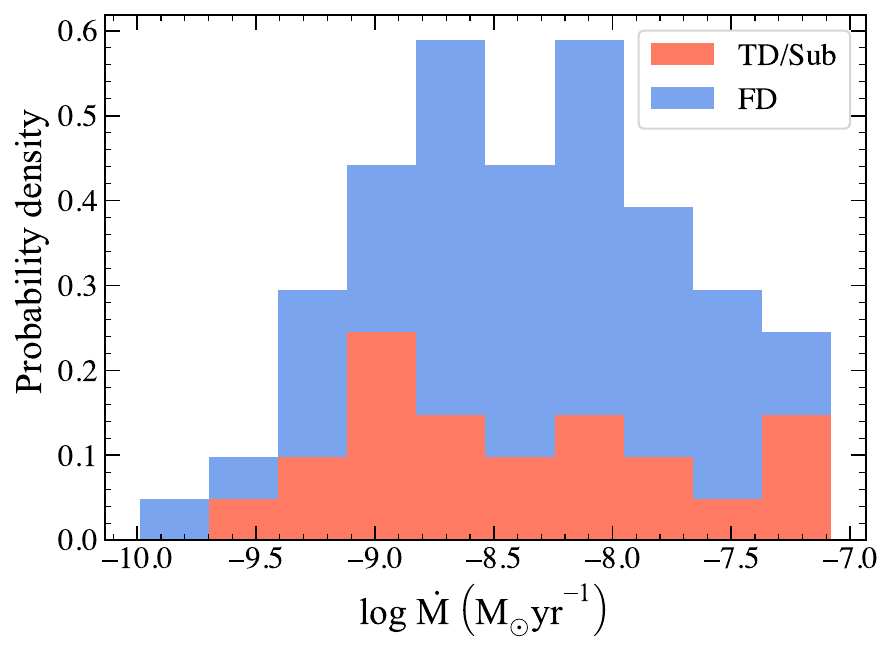}
\caption{Distribution of abundance by number, colored according to the presence (red, TD/SubS) or absence of structure in the disk (blue, FD )  in each system.}
\label{fig: mdot-struc}
\end{figure}

We speculate that a possible explanation is that the inner and outer disks are decoupled and that the mass accretion rate mostly measures properties of the inner disk, while refractory depletion mostly reflects phenomena in the outer disks, related to the presence of structure and forming planets. To support this idea, we performed a two-sample KS test comparing the underlying mass accretion rate distributions of disks with and without structure (cavities or substructures), obtaining a statistic of 0.21 and a $p$-value of 0.46. This result indicates that it is unlikely that the underlying \mdot\ distributions of structured and smooth disks differ from each other (cf. Fig. \ref{fig: mdot-struc}).

The dichotomy between the inner and the outer disks has been explored before by proposing the existence of a mass reservoir in the inner disk, which would feed the accretion flows onto the star, while the outer disk evolves and may form planets. 
This mass reservoir could keep a relatively high mass accretion rate onto the star, even when giant planets are forming in the disk and draining the gas flow from the outer disks
\citep{zhu_transitional_2011}; this, in turn, could explain the lack of massive disks accreting at very low accretion rates expected from planet population synthesis models, although the extent of the effects of accreting giant planets in the host star accretion rates is still debated \citep{manara_constraining_2019,bergezcasalou_influence_2020, bergezcasalou_simulations_2023}.
This mass reservoir could also reconcile the inner disk surface with that expected in the minimum mass solar nebula \citep{hartmann_how_2018}, and could explain the strong accretors at old ages \citep{ingleby_evolution_2014}.

\section{Summary and conclusions} \label{sec:con}

We used magnetospheric accretion models to analyze the VLT X-shooter spectra of 70 CTTS from the star-forming regions of Chamaeleon I, Lupus, and Orion OB1b. We calculated Ca abundances relative to hydrogen for the sample and explored the connection between our results and stellar and disk properties from the literature. Here we summarize the main results: 
\begin{enumerate}
    \item We find a wide range of Ca abundances, with Ca depletion present in all three star-forming regions. The overall distribution of Ca abundances is skewed toward low [Ca/H] values (high values of depletion), with 57\% of the sample having $[Ca/H] < -0.30$ relative to the solar. This indicates that refractory depletion is a very common result in CTTS, most likely caused by dust/disk evolution.
    \item The results of the two-sample KS test indicate that it is unlikely that the underlying abundance distributions between ChaI~-~Lup and ChaI~-~Ori 1b differ from each other, while the underlying distributions of Lup and Ori 1b are likely different from each other. This could hint at the difference in abundance distribution with age; however, given the small size of the Ori 1b sample, we cannot rule out the possibility that the underlying distributions of three star-forming regions in this work are similar.
    \item All transitional disks are depleted in calcium with $[Ca/H] < -0.15$.
    We find disks with known substructures in a wide range of abundance values, with CR Cha and Sz68 being the only structured disks with higher abundance. In both cases the inner disk is still connected to the outer parts, and we suggest that there is no trapping mechanism diminishing the flow of rocky material to the central stars in these cases.
    \item We find $\sim 60\%$ of the full disks also show significant Ca depletion, with $[Ca/H] < -0.30$. However, we cannot rule out the possibility of hidden substructures and/or cavities in these sub-samples. This is especially true for ChaI, where most of the stars were observed at a resolution of 0.6", resolving only cavities $\geq$50 au \citep{pascucci_steeper_2016}. However, we still find full disks with high depletion values in Lupus, suggesting that this result is not only due to a detection bias. 
    \item For the ChaI and Lup subsamples, no correlation is found between Ca abundance and stellar mass nor disk dust mass.  
    \item We find an anti-correlation between $[Ca/H]$ and the mass accretion rate, with a slope of $\sim\ -1$. Since the mass accretion is directly related to the abundance of gas, mostly represented by H, the anticorrelation would seem to imply that the abundance of solids reaching the inner disk is approximately constant, while \halpha\ scales with mass accretion rate, assuming that the Ca abundance is a proxy for the solids. A possibility for this to happen is that there is an additional contribution to the infalling gas from an inner disk reservoir such as a dead zone, implying a decoupling of the inner and outer disk.
    \item  Our results in terms of depletion and timescales for depletion are qualitatively consistent with the expectations of dust growth and radial drift, including partitioning of elements \citep{huhn_how_2023}, and constitute direct proof that radial drift of solids locked in pebbles takes place in disks.
\end{enumerate}

\vspace{10pt}
We thank Bertram Bitsch, Lee Hartmann, and Zhaohuan Zhu for enlightening conversations and suggestions.
This work has been partially supported by NASA grant 80NSSC24K0151.

Based on observations collected at the European Southern Observatory under ESO programmes 106.20Z8.002, 084.C-0269(A), 085.C-0238(A), 086.C-0173(A), 087.C-0244(A), 089.C-0143(A), 095.C-0134(A), 097.C-0349(A), 095.C-0378(A), 084.C-1095(A),
funded by the European Union (ERC, WANDA, 101039452). Views and opinions expressed are, however, those of the author(s) only and do not necessarily reflect those of the European Union or the European Research Council Executive Agency. Neither the European Union nor the granting authority can be held responsible for them.

JMA acknowledges financial support from PRIN-MUR 2022 20228JPA3A “The path to star and planet formation in the JWST era (PATH)” funded by NextGeneration EU and by INAF-GoG 2022 “NIR-dark Accretion Outbursts in Massive Young stellar objects (NAOMY)”. 
JMA and CFM acknowledge financial support from Large Grant INAF 2022 “YSOs Outflows, Disks and Accretion: towards a global framework for the evolution of planet forming systems (YODA)”.

The authors also acknowledge the referee for their useful comments, which improved the original manuscript.

\software{\texttt{Astropy} \citep[][]{astropy_collaboration_astropy_2013,astropy_collaboration_astropy_2018}, \texttt{PyAstronomy} \citep[][]{czesla_pya_2019},  \texttt{Eniric} \citep[][]{neal_jason-nealeniric_2019},
\texttt{Scipy} \citep[][]{virtanen_scipy_2020}, \texttt{Linmix}
\citep{linmixgithub}.
}
\bibliography{bib.bib}{}
\bibliographystyle{aasjournal}

\appendix
\restartappendixnumbering
\renewcommand\theHtable{Appendix.\thetable}

\section{Tables}
This appendix contains tables with TTSs properties used in the analysis of this work.

\startlongtable
\begin{deluxetable*}{lccccccccccc}
\tablecaption{Stellar parameters for our CTTSs sample \label{tab:ctts}}
\tabletypesize{\scriptsize}
\tablehead{\colhead{Name} & \colhead{2MASS} & \colhead{SpT} & \colhead{T$_{\rm eff}$(K)} & \colhead{A$_{\text{V}}$} & \colhead{M($M_{\sun}$)} & \colhead{R($R_{\sun}$)} & \colhead{log $\dot{\text{M}}$} & \colhead{d(pc)} & \colhead{$\rm M_{dust}^1$($\rm M_{\oplus}$)} & \colhead{$\rm R_{68}^1(au)$} &  \colhead{Disk} }
\startdata
\cutinhead{ChaI$^2$}
CHX18N & J11114632-7620092 & K2 & 4900.0 & 0.8 & 1.25 & 1.41 & -8.09 &160.0 & 13.39 & 26.8254       & FD (r$^a$) \\  
CHXR 47 & J11103801-7732399 & K4 & 4590.0 & 3.9 & 1.32 & 2.18 & -8.12 &160.0  & 2.01 & ...      & FD (ur$^a$) \\  
CR Cha & J10590699-7701404 & K0 & 5110.0 & 1.3 & 1.77 & 2.3 & -8.71 &160.0  & 157.11 & 66.672    & SubS$^b$ \\  
CS Cha & J11022491-7733357 & K2 & 4900.0 & 0.8 & 1.4 & 1.67 & -8.29 &160.0  & 84.4 & 47.75       & TD$_c^c$ \\  
CW Cha & J11123092-7644241 & M0.5 & 3780.0 & 2.1 & 0.59 & 0.989 & -8.03 &160.0  & 4.76 & $<$25.2525      & FD (ur$^a$) \\ 
ESO-Ha-562 & J11080297-7738425 & M1 & 3705.0 & 3.4 & 0.56 & 0.841 & -9.24 &160.0  & 38.23 & ...     & FD (r$^a$) \\  
... & J11085367-7521359 & M1 & 3705.0 & 1.5 & 0.51 & 1.06 & -8.15 &160.0  & 9.25 & 28.5      & FD (r$^a$) \\ 
...  & J11432669-7804454 & M5.5 & 3060.0 & 0.4 & 0.14 & 1.07 & -8.71 &160.0  & $<$0.56 & ...      & FD (r$^a$) \\ 
Sz Cha & J10581677-7717170 & K2 & 4900.0 & 1.3 & 1.31 & 1.5 & -7.82 &160.0  & 116 & 118.42     & TD$_c^c$ \\  
Sz18 & J11071915-7603048 & M2 & 3560.0 & 1.3 & 0.38 & 1.34 & -8.7 &160.0  & ... & ...       & TD$_u^d$ \\  
Sz19 & J11072074-7738073 & K0 & 5110.0 & 1.5 & 2.08 & 2.88 & -7.63 &160.0  & 9.75 & $<$17.0091       & FD (r$^a$) \\  
Sz22 & J11075792-7738449 & K5 & 4350.0 & 3.2 & 1.01 & 1.26 & -8.34 &160.0  & 7.42 & ...      &  FD (r$^a$) \\  
Sz27 & J11083905-7716042 & K7 & 4060.0 & 2.9 & 0.8 & 1.16 & -8.86 &160.0  & 5.27 & $<$17.082       & TD$_u^d$ \\  
Sz32 & J11095340-7634255 & K7 & 4060.0 & 4.3 & 0.78 & 1.4 & -7.08 &160.0  & 28.46 & 68.76      &  FD (r$^a$) \\  
Sz33 & J11095407-7629253 & M1 & 3705.0 & 1.8 & 0.56 & 0.805 & -9.35 &160.0  & 11.4 & ...       &  FD (r$^a$) \\  
Sz37 & J11104959-7717517 & M2 & 3560.0 & 2.7 & 0.41 & 1.02 & -7.82 &160.0  & 21.07 & 39.2049 &  FD (r$^a$) \\  
Sz45 & J11173700-7704381 & M0.5 & 3780.0 & 0.7 & 0.51 & 1.51 & -8.09 &160.0  & 10.44 & 30.2208      & TD$_u^d$ \\  \
T10 & J11004022-7619280 & M4 & 3270.0 & 1.1 & 0.23 & 0.985 & -9.22 &160.0  & 26.92 & 48.2525      & FD (r$^a$) \\  
T12 & J11025504-7721508 & M4.5 & 3200.0 & 0.8 & 0.19 & 1.26 & -8.7 &160.0  & 0.41 & $<$42.366     & TD$_u^e$ \\  
T16 & J11045701-7715569 & M3 & 3415.0 & 4.9 & 0.29 & 1.51 & -7.8 &160.0  & 1.43 & $<$23.2308      & FD (ur$^a$) \\  
T23 & J11065906-7718535 & M4.5 & 3200.0 & 1.7 & 0.21 & 1.84 & -8.11 &160.0  & 8.84 & 22.5024       & FD (r$^a$) \\  
T24 & J11071206-7632232 & M0 & 3850.0 & 1.5 & 0.58 & 1.42 & -8.49 &160.0  & 2.52 & 42.6514     & FD (ur$^a$) \\  
T27 & J11072825-7652118 & M3 & 3415.0 & 1.2 & 0.29 & 1.67 & -8.36 &160.0  & 0.81 & ...     & FD (nd$^a$) \\  
T28 & J11074366-7739411 & M1 & 3705.0 & 2.8 & 0.48 & 1.33 & -7.92 &160.0  & 40.22 & 32.3408      & FD (r$^a$) \\  
T3 & J10555973-7724399 & K7 & 4060.0 & 2.6 & 0.77 & 0.858 & -8.61 &160.0  & 2.28 & 20.1828     & FD (r$^a$) \\  
T3 B & J10555973-7724399 & M3 & 3415.0 & 1.3 & 0.29 & 1.25 & -8.43 &160.0  & 2.28 & 20.1828       & ... \\  
T30 & J11075809-7742413 & M3 & 3415.0 & 3.8 & 0.3 & 1.14 & -8.31 &160.0  & 2.55 & $<$15.0016     & FD (ur$^a$) \\  
T33 B & J11081509-7733531 & K0 & 5110.0 & 2.7 & 1.0 & 1.06 & -8.69  &160.0 & 78.27 & 63.03      & ... \\  
T38 & J11085464-7702129 & M0.5 & 3780.0 & 1.9 & 0.63 & 0.841 & -9.3 &160.0  & 1.49 & $<$24.83       & FD (ur$^a$) \\  
T4 & J10563044-7711393 & K7 & 4060.0 & 0.5 & 0.78 & 1.33 & -9.41 &160.0  & $<$1.33 & 139.1621       & TD$_c^c$ \\  
T40 & J11092379-7623207 & M0.5 & 3780.0 & 1.2 & 0.49 & 1.73 & -7.33 &160.0  & 46.58 & 32.4904      & FD (r$^a$ ) \\  
T46 & J11100704-7629376 & K7 & 4060.0 & 1.2 & 0.75 & 1.47 & -8.7 &160.0  & 2.68 & ...         &  FD (ur$^a$) \\  
T48 & J11105333-7634319 & M3 & 3415.0 & 1.2 & 0.3 & 1.14 & -7.96 &160.0  & 12 & 30.9168       &  FD (r$^a$) \\  
T49 & J11113965-7620152 & M3.5 & 3340.0 & 1.0 & 0.25 & 1.61 & -7.41 &160.0  & 8.03 & $<$17.19     &  FD (ur$^a$) \\  
T52 & J11122772-7644223 & K0 & 5110.0 & 1.0 & 1.62 & 2.04 & -7.48 &160.0  & 22.49 & 23.0112     &  FD (r$^a$) \\  
TW Cha & J10590108-7722407 & K7 & 4060.0 & 0.8 & 0.79 & 1.25 & -8.86 &160.0  & 22.7 & 42.1176      &  FD (r$^a$) \\  
VW Cha & J11080148-7742288 & K7 & 4060.0 & 1.9 & 0.67 & 2.59 & -7.6  &160.0 & 16.59 & ...       &  FD (ur$^a$) \\  
\cutinhead{Lup$^{3}$}
RYLup & J15592838-4021513 & K2 & 4900.0 & 0.4 & 1.4 & 1.79 & -8.19 &150.0 & 64.6 & 99.54       & TD$_c^c$ \\  
SSTc2dJ160927 & J16092697-3836269 & M4.5 & 3200.0 & 2.2 & 0.19 & 1.1 & -7.93 &200.0 & 1.34 & ...      & FD (ur$^f$) \\ 
SSTc2dJ161243 & J16124373-3815031 & M1 & 3705.0 & 0.8 & 0.44 & 1.91 & -8.76 &160.0 & 9.04 & $<$11.1895      & FD (r$^f$) \\  
Sz110 & J16085157-3903177 & M4 & 3270.0 & 0.0 & 0.22 & 1.61 & -8.53 &200.0 & 5.49 & 15.749     & FD (r$^f$) \\  
Sz111 & J16085468-3937431 & M1 & 3705.0 & 0.0 & 0.47 & 1.4 & -9.12 &200.0 & 45.44 & 72.8502      & TD$_c^c$  \\  
Sz113 & J16085780-3902227 & M4.5 & 3197.0 & 1.0 & 0.2 & 0.83 & -8.87 &200.0 & 8.1 & 12.8424      & FD (r$^f$) \\  
Sz114 & J16090185-3905124 & M4.8 & 3175.0 & 0.3 & 0.21 & 1.82 & -8.96 &200.0 & 30.61 & 42.3252       & SubS$^g$ \\ 
Sz123A & J16105158-3853137 & M1 & 3705.0 & 1.25 & 0.51 & 1.1 & -8.86 &200.0 & $<$0.23 & ...      & TD$_c^c$ \\  
Sz123A & J16105158-3853137 & M1 & 3705.0 & 1.25 & 0.51 & 1.1 & -8.86 &200.0 & 12.7 & ...       & TD$_c^c$ \\  
Sz123B & J16105158-3853137 & M2 & 3560.0 & 0.0 & 0.46 & 0.58 & -9.99 &200.0 & 12.7 & ...       & FD (nd$^f$) \\  
Sz129 & J15591647-4157102 & K7 & 4060.0 & 0.9 & 0.79 & 1.23 & -8.4 &150.0 & 58.49 & 49.6403       & TD$_c^c$ \\  
Sz66 & J15392828-3446180 & M3 & 3415.0 & 1.0 & 0.29 & 1.29 & -8.54 &150.0 & 4.69 & $<$12.4736       & FD (ur$^f$) \\  
Sz68 & J15451286-3417305 & K2 & 4900.0 & 1.0 & 1.4 & 3.14 & -8.24 &150.0 & 49.8 & ...     & SubS$^g$ \\  
Sz69 & J15451741-3418283 & M4.5 & 3197.0 & 0.0 & 0.2 & 0.97 & -9.51 &150.0 & 5.63 & $<$9.1536       & FD (r$^f$) \\  
Sz71 & J15464473-3430354 & M1.5 & 3632.0 & 0.5 & 0.42 & 1.43 & -9.06 &150.0 & 50.06 & 69.84   & SubS$^g$ \\  
Sz72 & J15475062-3528353 & M2 & 3560.0 & 0.75 & 0.37 & 1.29 & -8.65 &150.0 & 4.09 & $<$10.9697     & FD (r$^f$) \\  
Sz73 & J15475693-3514346 & K7 & 4060.0 & 3.5 & 0.79 & 1.35 & -8.16 &150.0 & 8.11 & 31.564      & FD (r$^f$) \\  
Sz74 & J15480523-3515526 & M3.5 & 3342.0 & 1.5 & 0.3 & 3.13 & -7.87 &150.0 & 6.15 & ... & FD (r$^f$) \\  
Sz81A & J15555030-3801329 & M4.5 & 3200.0 & 0.0 & 0.19 & 1.54 & -8.98 &150.0 & 3.19 & ... & FD (ur$^f$) \\  
Sz83 & J15564230-3749154 & K7 & 4060.0 & 0.0 & 0.67 & 2.39 & -7.14 &150.0 & 125.22 & 47.4 & SubS$^g$ \\  
Sz88A & J16070061-3902194 & M0 & 3850.0 & 0.25 & 0.56 & 1.61 & -8.13 &150.0 & 2.78 & ... & TD$_c^c$ \\  
Sz88A & J16070061-3902194 & M0 & 3850.0 & 0.25 & 0.56 & 1.61 & -8.13 &150.0 & $<$0.23 & ...       & FD (r$^f$) \\  
Sz90 & J16071007-3911033 & K7 & 4060.0 & 1.8 & 0.73 & 1.64 & -8.64 &200.0 & 6.76 & 16.037 & FD (r$^f$) \\  
Sz91 & J16071159-3903475 & M1 & 3705.0 & 1.2 & 0.47 & 1.36 & -8.73 &200.0 & 7.27 & ... & TD$_c^c$ \\  
Sz98 & J16082249-3904464 & K7 & 4060.0 & 1.0 & 0.7 & 3.2 & -7.23 &200.0 & 75.85 & 128.1414      & SubS$^i$ \\ 
\cutinhead{OB1b$^{4}$}
CVSO104 & J05320638-0111000 & M0.5 & 3700.0 & 0.05 & 0.37 & 1.64 & -7.94 & 360.7 & ... & ...  & TD$_u^h$ \\  
CVSO107 & J05322578-0036533 & K6 & 4020.0 & 1.16 & 0.53 & 1.97 & -7.32 & 330.4 & ... & ... & TD$_u^h$ \\  
CVSO109 & J05323265-0113461 & M0.5 & 3799.0 & 0.1 & 0.46 & 2.21 & -7.49 & 400.0 & ... & ... & TD$_u^{e,h}$ \\  
CVSO146 & J05354600-0057522 & M2 & 3490.0 & 0.28 & 0.86 & 1.25 & -8.28 & 332.0 & ... & ...     & ... \\  
CVSO165 & J05390257-0120323 & K6 & 4040.0 & 0.2 & 0.84 & 2.02 & -9.1 & 400.0 & ... & ...    & ... \\   
CVSO176 & J05402414-0031213 & M1 & 3849.0 & 1.44 & 0.25 & 3.26 & -7.38 & 302.4 & ... & ...       & ... \\  
CVSO58 & J05292326-0125153 & K7 & 3970.0 & 1.39 & 0.81 & 1.05 & -7.99 & 349.0 & ... & ...      & ... \\  
CVSO90 & J05312062-0049197 & M0.5 & 3700.0 & 0.92 & 0.62 & 0.84 & -7.9 & 338.7 & ... & ... & ...  
\enddata
\tablenotetext{}{SSTc2dJ1609 and SSTc2dJ1612 full name are  SSTc2dJ160927.0-383628 and SSTc2dJ161243.8-381503, respectively.}
\tablerefs{ Disk parameters from: {1} \cite{manara_demographics_2023} and references therein (see \S \ref{subsec:ctts}). Stellar properties from: {2} \cite[ChaI:][]{manara_x-shooter_2016,manara_x-shooter_2017},  {3} \cite[Lup:][]{alcala_x-shooter_2014,alcala_x-shooter_2017}, {4} \cite[OB1b:][]{manara_penellope_2021,pittman_towards_2022} {\bf. Typical uncertainties in mass accretion rate are 0.35dex and 0.42dex for stars of the ChaI and Lup regions, respectively. For the stars of Ori 1b, the mass accretion rates have typical uncertainties of 0.016dex, except CVSO109 and CVSO165, which have typical errors of 0.45dex. }  }
\tablenotetext{{\bf Disk Morphology:}}{\\ 
{\bf FD:} Full disks.
\begin{itemize}
\vspace{-5pt}
\itemsep0em 
\item {\bf r:} Disk resolved with millimeter imaging, no cavities or substructures detected (a: \cite{pascucci_steeper_2016}, f: \cite{ansdell_alma_2016})
\item {\bf ur:} Disk unresolved with millimeter imaging (a: \cite{pascucci_steeper_2016}, f: \cite{ansdell_alma_2016})
\item {\bf nd:} Non-detection of the disk with millimeter imaging (a: \cite{pascucci_steeper_2016}, f: \cite{ansdell_alma_2016})
\end{itemize}
\vspace{-5pt}
{\bf SubS:} Substrucres detected with milimiter imaging; e.g., ring in the outer disk, spiral arms
(b: \cite{kim_detection_2020}, g: \cite{andrews_disk_2018}, i: \cite{gasman_sz98_2023})\\
{\bf TD:} Transitional Disks 
\begin{itemize}
\vspace{-5pt}
\itemsep0em 
\item {\bf TD$_c$:} Transitional disks, cavities confirmed by millimeter imaging \citep[c:][]{van_der_Marel_transition_2023}.
\item {\bf TD$_u$:} Transitional disks, cavities predicted by SED
(d: \cite{manoj_spitzer_2011}, h: \cite{mauco_herschel_2018}) or by Ly$\alpha$ line profile \citep[e:][]{arulanantham_lya_2023}, but not confirmed by millimeter imaging.
\end{itemize}
}
\end{deluxetable*}

\begin{deluxetable}{lccccccccc}[h]
\tablecaption{Names, stellar parameters and line fluxes for our WTTSs Sample
\label{tab:wtts}}
\tablehead{\colhead{Name} & \colhead{SpT} & \colhead{d(pc)} & \colhead{T$_{\rm eff}$(K)}  & \colhead{L($L_{\sun}$)} & \colhead{R($R_{\sun}$)} & \colhead{F(\halpha)} & \colhead{$\sigma$(\halpha)} & \colhead{F(\CaIIk)} & \colhead{$\sigma$(\CaIIk)}}
\startdata
HBC 407               & K0   & 140.0 & 5110.0 & 0.36 & 0.76 & 24.1 & 11.5 & 29.2 & 14.9 \\
PZ99 J160843.4-260216 & K0.5 & 145.0 & 5050.0 & 1.38 & 1.53 & 221.0 & 65.7 & 456.0 & 123.0 \\
RX J1515.8-3331       & K0.5 & 150.0 & 5050.0 & 1.25 & 1.46 & 114.0 & 33.2 & 135.0 & 36.4 \\
PZ99 J160550.5-253313 & K1   & 145.0 & 5000.0 & 0.98 & 1.32 & 156.0 & 47.0 & 260.0 & 69.4 \\
RX J0438.6+1546       & K2   & 140.0 & 4900.0 & 0.95 & 1.35 & 186.0 & 32.9 & 151.0 & 25.7 \\
RX J1547.7-4018       & K3   & 150.0 & 4730.0 & 0.83 & 1.36 & 70.0 & 23.2 & 109.0 & 20.0 \\
RX J1538.6-3916       & K4   & 150.0 & 4590.0 & 0.61 & 1.23 & 82.1 & 20.3 & 108.0 & 16.9 \\
TWA9A                 & K5   & 68.0  & 4350.0 & 0.25 & 0.872 & 205.0 & 18.1 & 125.0 & 7.31 \\
RX J1543.1-3920       & K6   & 150.0 & 4205.0 & 0.40 & 1.19 & 71.1 & 8.91 & 58.0 & 3.94 \\
RX J1540.73756       & K6   & 150.0 & 4205.0 & 0.39 & 1.18 & 59.8 & 9.36 & 54.9 & 3.75 \\
SO879                 & K7   & 360.0 & 4060.0 & 0.51 & 1.45 & 24.5 & 1.88 & 13.0 & 0.526 \\
TWA25                 & M0   & 54.0  & 3850.0 & 0.25 & 1.11 & 528.0 & 31.0 & 224.0 & 8.41 \\
TWA14                 & M0.5 & 96.0  & 3780.0 & 0.15 & 0.897 & 157.0 & 4.81 & 37.9 & 1.07 \\
TWA13B                & M1   & 59.0  & 3705.0 & 0.20 & 1.08 & 302.0 & 20.3 & 144.0 & 5.07 \\
TWA7                  & M2   & 28.0  & 3415.0 & 0.07 & 0.769 & 476.0 & 17.5 & 136.0 & 3.36 \\
TWA2A                 & M2   & 47.0  & 3560.0 & 0.33 & 1.51 & 570.0 & 41.1 & 279.0 & 10.3 \\
TWA15B                & M2   & 111.0 & 3415.0 & 0.11 & 0.946 & 69.9 & 1.38 & 15.0 & 0.267 \\
TWA9B                 & M3   & 68.0  & 3415.0 & 0.07 & 0.743 & 72.5 & 2.59 & 21.5 & 0.641 \\
TWA15A                & M3.5 & 111.0 & 3340.0 & 0.11 & 1.0 & 103.0 & 1.51 & 18.8 & 0.292 \\
Sz94                  & M4   & 200.0 & 3270.0 & 0.17 & 1.3 & 27.3 & 0.84 & 7.12 & 0.197 \\
SO797                 & M4.5 & 360.0 & 3200.0 & 0.06 & 0.763 & 1.79 & 0.05 & 0.302 & 0.007 \\
SO641                 & M5   & 360.0 & 3125.0 & 0.03 & 0.586 & 1.21 & 0.03 & 0.169 & 0.004 \\
Par-Lup3-2            & M5   & 200.0 & 3125.0 & 0.18 & 1.44 & 10.7 & 0.38 & 1.3 & 0.079 \\
SO999                 & M5.5 & 360.0 & 3060.0 & 0.05 & 0.815 & 2.03 & 0.04 & 0.211 & 0.005 \\
SO925                 & M5.5 & 360.0 & 3060.0 & 0.03 & 0.57 & 0.83 & 0.02 & 0.085 & 0.002 \\
LM 717                & M6.5 & 160.0 & 2935.0 & 0.02 & 0.516 & 1.35 & 0.03 & 0.075 & 0.004
\enddata
\tablecomments{Fluxes in units of $10^{-15} \rm erg \ s^{-1} \ cm^{-2}$}
\tablerefs{ Stellar properties from \cite{manara_x-shooter_2013, manara_extensive_2017}, fluxes from \cite{micolta_ca_2023}. }
\end{deluxetable}

\startlongtable
\begin{deluxetable*}{lcccccccccc}
\label{tab:results}
\tablecaption{Designations, line fluxes and Ca abundances for the T Tauri stars in our sample}
\tablehead{\colhead{Name} & \colhead{2MASS} & \colhead{F(\halpha)} & \colhead{$\sigma$(\halpha)} & \colhead{F(\CaIIk)} & \colhead{$\sigma$(\CaIIk)} & 
\colhead{$\rm X$$^a$} & \colhead{$\sigma$($\rm X$)} &
\colhead{$[\rm{Ca/H}]$} & \colhead{$\sigma$($[\rm{Ca/H}]$)}}
\startdata
\cutinhead{ChaI$^1$}
CHX18N & J11114632-7620092 & 172.0 & 19.5 & 19.8 & 3.99 & 0.42 & 0.36 & -0.38 & 0.27 \\
CHXR 47 & J11103801-7732399 & 77.5 & 34.0 & 43.4 & 10.5 & 0.82 & 0.3 & -0.09 & 0.13 \\
CR Cha & J10590699-7701404 & 1140.0 & 70.9 & 94.6 & 25.8 & 0.83 & 0.34 & -0.08 & 0.15 \\
CS Cha & J11022491-7733357 & 709.0 & 34.3 & 52.5 & 10.2 & 0.45 & 0.38 & -0.34 & 0.27 \\
CW Cha & J11123092-7644241 & 237.0 & 6.4 & 43.3 & 5.17 & 0.27 & 0.2 & -0.57 & 0.24 \\
ESO-Ha-562 & J11080297-7738425 & 89.6 & 1.66 & 13.6 & 0.34 & 1.0* & ... & 0.0* & ... \\
J11085367-7521359 & J11085367-7521359 & 176.0 & 4.83 & 20.1 & 3.18 & 0.23 & 0.2 & -0.64 & 0.28 \\
J11432669-7804454 & J11432669-7804454 & 45.3 & 0.77 & 2.35 & 0.13 & 0.22 & 0.2 & -0.65 & 0.27 \\
Sz Cha & J10581677-7717170 & 301.0 & 33.0 & 26.0 & 11.8 & 0.22 & 0.18 & -0.65 & 0.25 \\
Sz18 & J11071915-7603048 & 45.8 & 2.86 & 2.99 & 0.2 & 0.52 & 0.43 & -0.28 & 0.26 \\
Sz19 & J11072074-7738073 & 1180.0 & 105.0 & 144.0 & 78.4 & 0.22 & 0.18 & -0.66 & 0.26 \\
Sz22 & J11075792-7738449 & 273.0 & 8.93 & 75.7 & 5.47 & 0.76 & 0.37 & -0.12 & 0.17 \\
Sz27 & J11083905-7716042 & 191.0 & 6.13 & 4.62 & 0.47 & 0.24 & 0.2 & -0.63 & 0.26 \\
Sz32 & J11095340-7634255 & 2350.0 & 39.8 & 1200.0 & 79.6 & 0.2 & 0.13 & -0.71 & 0.21 \\
Sz33 & J11095407-7629253 & 25.5 & 1.36 & 2.06 & 0.17 & 0.98 & 0.03 & -0.01 & 0.01 \\
Sz37 & J11104959-7717517 & 274.0 & 6.47 & 117.0 & 5.55 & 0.48 & 0.27 & -0.32 & 0.19 \\
Sz45 & J11173700-7704381 & 272.0 & 6.35 & 16.2 & 2.08 & 0.11 & 0.11 & -0.96 & 0.3 \\
T10 & J11004022-7619280 & 41.7 & 0.87 & 1.31 & 0.1 & 0.62 & 0.42 & -0.21 & 0.22 \\
T12 & J11025504-7721508 & 33.6 & 0.77 & 1.79 & 0.14 & 0.22 & 0.2 & -0.65 & 0.27 \\
T16 & J11045701-7715569 & 27.0 & 3.38 & 18.4 & 1.74 & 0.64 & 0.31 & -0.19 & 0.17 \\
T23 & J11065906-7718535 & 140.0 & 2.29 & 29.7 & 0.61 & 0.21 & 0.18 & -0.69 & 0.28 \\
T24 & J11071206-7632232 & 38.6 & 6.13 & 5.74 & 0.44 & 0.63 & 0.41 & -0.2 & 0.22 \\
T27 & J11072825-7652118 & 68.1 & 3.33 & 10.3 & 0.73 & 0.36 & 0.35 & -0.45 & 0.3 \\
T28 & J11074366-7739411 & 312.0 & 7.6 & 17.0 & 1.22 & 0.07 & 0.04 & -1.17 & 0.2 \\
T3 & J10555973-7724399 & 137.0 & 4.07 & 15.1 & 1.73 & 0.64 & 0.41 & -0.19 & 0.21 \\
T3 B & J10555973-7724399 & 32.9 & 1.82 & 6.47 & 0.58 & 0.54 & 0.45 & -0.27 & 0.26 \\
T30 & J11075809-7742413 & 126.0 & 2.7 & 27.7 & 0.94 & 0.45 & 0.4 & -0.35 & 0.28 \\
T33 B & J11081509-7733531 & 276.0 & 15.9 & 47.0 & 11.6 & 0.98 & 0.03 & -0.01 & 0.02 \\
T38 & J11085464-7702129 & 39.0 & 1.95 & 3.65 & 0.33 & 0.98 & 0.03 & -0.01 & 0.01 \\
T4 & J10563044-7711393 & 25.0 & 3.38 & 5.28 & 0.24 & ... & ... & ... & ... \\
T40 & J11092379-7623207 & 471.0 & 15.7 & 83.5 & 13.3 & 0.08 & 0.05 & -1.12 & 0.21 \\
T46 & J11100704-7629376 & 206.0 & 9.48 & 5.49 & 0.7 & 0.2 & 0.17 & -0.69 & 0.27 \\
T48 & J11105333-7634319 & 396.0 & 3.44 & 91.0 & 1.87 & 0.24 & 0.19 & -0.62 & 0.26 \\
T49 & J11113965-7620152 & 237.0 & 5.29 & 9.3 & 4.74 & 0.02 & 0.02 & -1.66 & 0.28 \\
T52 & J11122772-7644223 & 3010.0 & 101.0 & 919.0 & 57.1 & 0.35 & 0.24 & -0.45 & 0.22 \\
TW Cha & J10590108-7722407 & 122.0 & 5.21 & 15.9 & 0.96 & 0.83 & 0.32 & -0.08 & 0.14 \\
VW Cha & J11080148-7742288 & 1130.0 & 32.0 & 59.0 & 4.3 & 0.05 & 0.04 & -1.34 & 0.27 \\
\cutinhead{Lup}
RYLup & J15592838-4021513 & 239.0 & 39.3 & 44.5 & 13.5 & 0.69 & 0.38 & -0.16 & 0.19 \\
SSTc2dJ160927 & J16092697-3836269 & 32.0 & 0.96 & 25.0 & 0.73 & 0.61 & 0.35 & -0.21 & 0.2 \\
SSTc2dJ161243 & J16124373-3815031 & 48.2 & 4.11 & 5.37 & 0.32 & 0.73 & 0.4 & -0.13 & 0.19 \\
Sz110 & J16085157-3903177 & 47.4 & 1.56 & 4.72 & 0.4 & 0.38 & 0.38 & -0.42 & 0.3 \\
Sz111 & J16085468-3937431 & 114.0 & 2.69 & 1.33 & 0.17 & 0.22 & 0.19 & -0.66 & 0.27 \\
Sz113 & J16085780-3902227 & 26.0 & 0.26 & 17.4 & 0.19 & & 1.0* & ... & 0.0* & ... \\
Sz114 & J16090185-3905124 & 28.3 & 0.84 & 2.3 & 0.09 & 0.52 & 0.44 & -0.29 & 0.27 \\
Sz118 & J16094864-3911169 & 64.1 & 7.82 & 24.5 & 1.79 & ... & ... & ... & ... \\
Sz123A & J16105158-3853137 & 197.0 & 2.21 & 8.35 & 0.45 & 0.49 & 0.46 & -0.31 & 0.28 \\
Sz123B & J16105158-3853137 & 31.1 & 0.38 & 0.62 & 0.03 & & 1.0* & ... & 0.0* & ... \\
Sz129 & J15591647-4157102 & 113.0 & 7.48 & 19.2 & 2.17 & 0.59 & 0.43 & -0.23 & 0.24 \\
Sz66 & J15392828-3446180 & 66.0 & 2.12 & 6.07 & 0.44 & 0.36 & 0.36 & -0.44 & 0.3 \\
Sz68 & J15451286-3417305 & 621.0 & 127.0 & 114.0 & 36.5 & 0.73 & 0.38 & -0.14 & 0.18 \\
Sz69 & J15451741-3418283 & 73.4 & 0.46 & 5.27 & 0.06 & 0.88 & 0.29 & -0.05 & 0.12 \\
Sz71 & J15464473-3430354 & 118.0 & 2.88 & 4.15 & 0.21 & 0.58 & 0.44 & -0.24 & 0.25 \\
Sz72 & J15475062-3528353 & 178.0 & 2.81 & 56.3 & 0.55 & 0.88 & 0.29 & -0.05 & 0.12 \\
Sz73 & J15475693-3514346 & 272.0 & 8.26 & 72.0 & 2.75 & 0.54 & 0.42 & -0.27 & 0.25 \\
Sz74 & J15480523-3515526 & 111.0 & 7.94 & 22.3 & 1.12 & 0.19 & 0.17 & -0.71 & 0.27 \\
Sz81A & J15555030-3801329 & 32.4 & 0.89 & 1.45 & 0.05 & 0.32 & 0.37 & -0.5 & 0.33 \\
Sz83 & J15564230-3749154 & 1700.0 & 28.4 & 593.0 & 16.8 & 0.13 & 0.12 & -0.88 & 0.28 \\
Sz84 & J15580252-3736026 & 51.2 & 0.9 & 0.51 & 0.06 & 0.15 & 0.18 & -0.81 & 0.34 \\
Sz88A & J16070061-3902194 & 242.0 & 4.99 & 45.2 & 1.78 & 0.36 & 0.37 & -0.44 & 0.3 \\
Sz90 & J16071007-3911033 & 60.8 & 6.53 & 12.0 & 1.07 & 0.82 & 0.37 & -0.09 & 0.16 \\
Sz91 & J16071159-3903475 & 171.0 & 2.3 & 2.72 & 0.32 & 0.14 & 0.15 & -0.86 & 0.33 \\
Sz98 & J16082249-3904464 & 425.0 & 27.9 & 58.0 & 5.74 & 0.05 & 0.04 & -1.3 & 0.25 \\
\cutinhead{OB1b}
CVSO104 & J05320638-0111000 & 34.0 & 0.73 & 4.11 & 0.15 & 0.16 & 0.12 & -0.79 & 0.23 \\
CVSO107 & J05322578-0036533 & 30.7 & 2.24 & 15.1 & 1.13 & 0.28 & 0.12 & -0.55 & 0.15 \\
CVSO109 & J05323265-0113461 & 58.4 & 2.4 & 10.5 & 0.78 & 0.12 & 0.11 & -0.93 & 0.29 \\
CVSO146 & J05354600-0057522 & 44.9 & 2.84 & 3.36 & 0.22 & 0.21 & 0.13 & -0.67 & 0.21 \\
CVSO165 & J05390257-0120323 & 37.4 & 2.96 & 9.38 & 0.43 & ... & ... & ... & ... \\
CVSO176 & J05402414-0031213 & 64.4 & 1.56 & 3.1 & 0.55 & 0.03 & 0.02 & -1.56 & 0.25 \\
CVSO58 & J05292326-0125153 & 54.0 & 2.38 & 11.0 & 1.05 & 0.3 & 0.16 & -0.53 & 0.19 \\
CVSO90 & J05312062-0049197 & 105.0 & 1.25 & 35.9 & 0.8 & 0.4 & 0.2 & -0.4 & 0.18
\enddata
\tablenotetext{{\bf Notes.}}{\\ {\it a.} X refers to $\left(N_{Ca} / N_{H}\right) /\left(N_{Ca} / N_{H}\right)_{\odot}$, where N is the abundance by number.\\
{1} Fluxes from \cite{micolta_ca_2023}.\\
Fluxes in units of $10^{-14} \rm erg \ s^{-1} \ cm^{-2}$. \\
{*} CTTS fall at the edge of the models parameter space.}
\end{deluxetable*}

\end{document}